\documentclass[manuscript]{rmaa}
\usepackage{natbib}
\title{Physical mechanisms that shape the 
      morphologies of extragalactic
      jets}
\author{L. Zaninetti\altaffilmark{1}}

\altaffiltext{1}{Dipartimento di Fisica Generale, Via Pietro Giuria 1,  
10125 Torino,  Italy (zaninetti@ph.unito.it.)}
\shortauthor{Zaninetti}
\shorttitle {Jets}
\ReceivedDate{...............} 
\AcceptedDate{2} 
\SetYear{2000}
\resumen{Este documento describe ..
 } 

\abstract{
   In order to find a law characterising the decrease of
   velocity along
   a   jet,  five analytical
   methods are suggested.
    The first two simple models
examine the variation of velocity  in the presence
of Newton's or Stoke's  resistance.
The equation that represents the
conservation of the momentum along a pyramidal sector
is solved  from an analytical point of view (third model). 
The  application of the
conservation  of the total momentum flux allows us to deduce the
velocity of the galaxy as a function of  time 
for  classical velocities (fourth model) and relativistic velocities
(fifth model).
The variation of  velocity along the jet combined with an adequate
composition of jet precession  velocity~, rotational velocity  of
the galaxy~, and galaxy dispersion velocity in the cluster~ allows
us  to trace the geometrical pattern of the  head-tail radio
sources. 
 Application of the developed  theory/code to the  radio
galaxies NGC1265,NGC4061,NGC326, and  Cygnus~A gives  the central
galaxy's approximate dispersion velocities  in the direction
perpendicular to the jet. 
A
transition from head--tails to classical double radio  galaxies as
a function of  the increasing jet's mechanical power   is
introduced. 
}
 
\addkeyword{galaxies :jets} 
\addkeyword{ radio continuum : galaxies} 
\def\apj{ApJ\,}
\def\apjl{ApJ\,}
\def\aap{A\&A\,}
\def\mnras{MNRAS\,}
\def\aj{AJ}
\def\araa{ARA\&A}
\def\apjs{ApJS}
\def\POF{Physics of Fluids}

\begin{document}
\maketitle

\section{Introduction}

The extra-galactic radio sources are classified   on the basis of
the position  of the brightest radio emitting regions with respect
to the channel: FR-I have hot spots that are more distant from the
nucleus (typical example Cygnus~A) ,  FR-II radio galaxies have
emission uniformly  distributed along the channel (typical example
3C449). The physical parameter that governs this classification is
the radiated power: the more powerful radio galaxies being
classified as FR~II,see~\citet{fanaroff} for details. The
classification then becomes more complicated in the presence of
complex morphologies  and two  classifications are
introduced : NAT ( Narrow Angle Tail) and WAT ( Wide  Angle
Tail). But  a closer look at NGC~1265 reveals
that the angle subtended by the two channels is not so "narrow",
see Figure~1   of ~\citet{odea}; possibly,  in this case
a distinction  should be made 
between internal regions  of the channel where the
patterns are similar and the extended regions where conversely the
division in WAT and NAT is more evident.

For more details  on the classification scheme
the interested reader is advised to refer to
the book by  \citet{deyoung}.
The  X-ray observations  
of the Galactic black
hole XTE J1118+480
are 
consistent with
extended jets being the source of the hard X-ray flux.
In this hypothesis   the disc would
then simply represent a small solid angle as seen from the emission region,
see~\citet{Miller2002}. 
This point of view is not new and has been explored 
by~\citet{Mendoza_2005}  who suggested 
a  unified model for quasars and mu-quasars . 
In general, the shapes
seem  to be independent from (or weakly correlated to)
the cluster parameters, such as  physical position
in the cluster, number of galaxies, cluster richness,
cluster morphology and x-ray emission.
Some of the models that can explain
the bending of the jets will  now be briefly reviewed
\begin {itemize}
\item The {\it geometrical models}  started
        with
\citet{jaffe}
where two models were developed to explain
        the formation of "tailed" radio sources like 3C~129. 
        They continued inserting precession
        and  relativistic effects in SS433,
        see~\citet{hjellming}.
        An analytical model was presented for
        the evolution of a powerful double radio source
        on a small scale , see~\citet{alexander}.
\item  {\it Euler's equation model}, where  bending
        is produced by  the ram pressure $\rho_{ICM}V_j^2$
        of the intra-cluster medium, is  analysed
        by
~\citet{burns}
         in order to explain
        the structure of 1638+538~(4C53.57), and
        by
~\citet{venkatesan}
        in order to
        understand the behaviour  of NAT in
        poor  clusters of galaxies. Further on,
        a hydro-dynamical
        code was built   in order to explain the NAT sources,
        see ~\citet{norman}.
        From an observational
        point of view it is interesting to note that the
        velocities of the NAT galaxies are inadequate for producing
        the ram pressure necessary to bend the radio jets,
        see ~\citet{bliton}.

\item   The  {\it slingshot ejection model} consists in
        the ejection of black holes from the host galaxy:
        the bending
        is obtained from the oscillations about
        the center of the merged galaxy,
        see for example
~\citet{valtonen}.
\item
        The 3-dimensional magneto-hydrodynamic (MHD) simulations,
        based on the Sweeping
        Magnetic-Twist model
~\citet{nakamura},
  produce a
        wiggled structure of AGN radio jets.
\item 
The interaction between a jet  with a  
stratified cloud, see~\citet{Canto_and_Raga_1996},
or   with  a  spherically symmetric pressure stratification
, see~\citet{Raga_and_Canto_1996},
 can  lead
 to a final configuration in which the
jet has bored a hole through the cloud, 
or  large deflections are obtained.
\item A  jet can generate
an internal shock wave by  adopting  either a  non-relativistic 
framework , see \citet{Icke_jets1991} or a relativistic
framework, see \citet{Mendoza2002}.
\item 
From a relativistic point of view  \citet{Mendoza_Longair_2001}
showed
that bending relativistic jets is much more difficult than bending
non-relativistic ones. This is the reason why most FR-II radio galaxies
appear straight.

\end {itemize}

The already cited  models concerning the bending
leave a series of questions
unanswered or partially answered:
\begin {itemize}
\item Is it possible to
      deduce a law of motion in the  presence of viscosity
      in laminar and turbulent jets?

\item Is it possible to trace the jet velocity
      on the basis
      of physical parameters such  as
      linear density of  energy of
      the radio source and
      constant density of the IGM
      (Inter Galactic Medium)?

\item Is it possible to develop a consistent theory
      in which at a given  time the  jet opening
      angle  increases abruptly?
\item 
Is it possible to  include relativistic effects 
      in   turbulent jets
      that  generalise   the work made by 
      \citet{Canto_and_Raga_1996}
      and \citet{Raga_and_Canto_1996} 
      on the interaction between a jet and a stratified cloud    ?

\item Once  the main  jet trajectory is obtained, could
      other  kinematical  components be added
      such as  jet  precession,
      galaxy rotation and galaxy bulk velocity?

\item Does the developed theory  match the observed
      patterns traced by the radio sources?

\item Could  the transition FR-I~$\Rightarrow$~FR-II
      be simulated  by increasing the total  jet's
      luminosity?

\end {itemize}

In order  to answer these questions in Section~\ref{section_v} the
equations of motion were derived in the presence of Stoke's law of
resistance  ,  Newton's law of resistance  and turbulent eddy
viscosity. Some analytical computations on the momentum
conservation in an extra-galactic pyramidal sector  advancing in a
surrounding medium were developed in Section~\ref{sezione_a}. In
Section~\ref{twophase}  the theory of the two-phase beam was set
up. 
The theory  of the relativistic turbulent jets 
was developed in Section~\ref{twophaserelativistic}.
The  theory for the composition of the velocities  in a
head-tail radio source was developed in Section~\ref{sezione_c}.
The application of the developed  theory  to well specified  radio
sources such as  NGC1265,NGC4061, NGC326  and Cygnus A  was reported
in Section~\ref{sezione_d}.
A power transition is simulated in Section~\ref{sectransition} 
replacing  the energy with the total power in the law of motion.
The theory  of the  Kelvin-Helmholtz  instabilities was reviewed
and implemented  on  the radio-sources in Section~\ref{kh}.

\section{Viscosity models}

\label{section_v}
When a laminar jet   moves through  the IGM or the ISM a
retarding  drag
force $F_{drag}$ is applied. If $v$ is the instantaneous velocity
the simplest model which  is usually  considered is where
\begin {equation}
F_{drag} \propto  v ^n
\quad  ,
\end {equation}
with  $n$ as an integer.
Here the case  of  $n=1$ and $n=2$ is considered.
In classical mechanics $n=1$ is referred to  as Stoke's
law of resistance and   $n=2$ as Newton's
law of resistance.
We now consider  two
laminar jets and the turbulent case separately.

\subsection{Stoke's  behaviour}

\label{sec_stokes}
Consider a slice of the jet  with mass $m$.
The equation of motion is given by
\begin{equation}
m \frac {dv}{dt} = -b v
\end  {equation}
or
\begin{equation}
\frac {dv}{dt} = -B  v
\quad ,
\end  {equation}
where $ B= b/m$.
The  temporal behaviour of the velocity  turns out to be
\begin{equation}
v   =v_0 e ^{ -B t}
\label{sv}
\quad ,
\end  {equation}
where $v_0$  is the initial  velocity.
The distance at  the  time t  is
\begin{equation}
x  = \frac {v_0}{B} ( 1 - e ^{-Bt})
\quad  ,
\label{first}
\end  {equation}
and this can be considered the first law of motion.
The velocity space  dependence
is
\begin{equation}
v-v_0 = -Bx
\quad  .
\end {equation}
The velocity of the jet is allowed to vary between a maximum
value at the beginning (the inner region), 
$v=v_0=\frac{c}{c_f}$ with $c_f$ 
a number bigger than one ,
and a minimum value at the end (the outer region)
, for example $v=\frac{c}{10000}$.
In this    way  the coefficient $B$ can be found  
by  identifying x with the jet's length $x_1$
\begin{equation}
B = \frac { v_0 - c/10000} {x_1}
\quad .
\end{equation}
The  lifetime of the radiosource ,$t_{RS}^{Stoke}$,
can  be derived  from equation~(\ref{sv}) once
the numerical value of $B$ is known
\begin{equation}
t_{RS}^{Stoke} =   \frac {ln \frac{v_0 \times 10000}{c} }{B}
\quad  ,
\end{equation}
that  becomes
\begin{equation}
t_{RS}^{Stoke} =   \frac {ln \frac{ 10000}{c_f} }{B}
\quad  .
\end{equation}

\subsection{Newton's  behaviour}

\label{sec_newton}
The equation of motion is  now
\begin{equation}
m \frac {dv}{dt} = -a v^2
\quad ,
\end  {equation}
or
\begin{equation}
\frac {dv}{dt} = -  A v^2
\quad ,
\end  {equation}
where  $A= a/m$ .
The  temporal behaviour of the velocity  turns out to be
\begin{equation}
v   =\frac {v_0 }{1  +  v_0 A t }
\label{nv}
\quad ,
\end  {equation}
and the   distance
\begin{equation}
x  = \frac {ln (Av_0 t+1) }{A}
\label{second}
\quad  ,
\end  {equation}
which  can be considered the second law of motion.
The velocity space  dependence
is
\begin{equation}
x   = \frac {-ln~ \frac{v c_f}{ c}} {A}
\quad .
\end {equation}
The coefficient $A$ can be found
by inserting  
$\frac {v}{c}= 1/10000$  which  corresponds
to  $v=30~{\mathrm{km/sec}}$ while   $L$, the length of the jets,  is 
equal  to $x$ , later called $x_1$.

Here   the lifetime of the radiosource ,$t_{RS}^{Newton}$ , can  be
derived  from equation~(\ref{nv}) once the numerical value of $A$ is
known
\begin{equation}
t_{RS}^{Newton}  = \bigl (  { \frac { c}{c_f v}   - 1 } \bigr ) \frac {
c_f}{c A}
\quad  .
\end{equation}

\subsubsection{Parameters of the artificial viscosity from a sample of WAT}

\label{samplewat}
A sample of 7 radio--galaxies   classified
as Wide Angle Tails (WAT)
was  imaged
sensitively at high resolution by
~\citet{hardcastle}.
This  sample is here considered  as a source of data
that allows us  to derive the parameters of ~Section~\ref{sec_newton} and
Section~\ref{sec_stokes}, see  Table~\ref{A_B}.
The  parameters of the first two laws of motion are found once
the averaged velocity at the beginning of the jet (alias $c_f$ )
and at the end of the jet are fixed by observational arguments.
The spectroscopical measurements  are a good candidate
to find the jet's velocity 
but attention should be paid to the fact that 
at a given position x the velocity can vary
considerably 
 from the  boundary 
to the  center of the jet  , see formula~(\ref{ratiovel}).
 \begin{table}
 \caption[]{  A\lowercase{rtificial viscosity parameters of
               seven } WAT ; \\
            \lowercase {$c_f$ = 1.25  and velocity at the end v/c=1/10000}.
            T\lowercase{he  value of $c_f$ is chosen in
            order to have relativistic velocities in
            the initial stage (inner region)  of the jet   
            }.
           }
 \label{A_B}
 \[
 \begin{array}{lccccc}
 \hline
 \hline
 \noalign{\smallskip}
 Radio~name  & \mathrm {x_1~[{\mathrm{kpc}}]}  & t_{RS}^{Stoke} [10^6{\mathrm{
year}}]
                                    & B[1/{\mathrm{year}}]
                                    & t_{RS}^{Newton}  [10^6 {\mathrm{year}}]
                                    & A[1/\mathrm{pc}]
                                    \\
                                     \noalign{\smallskip}
 \hline
 \noalign{\smallskip}
0647+693 & 81 & 2.9   & 3.10~10^ {-6} & 2.8 10^2  & 1.1~10^ {-4}  \\ \noalign{\smallskip}
1231+674 & 35 & 1.25  & 7.13~10^{-6} & 1.2 10^2  & 2.56~10^{-4}   \\ \noalign{\smallskip}
1333+412 & 20 & 0.79  & 1.24~10^{-5} & 71        & 4.49~10^ {-4}  \\ \noalign{\smallskip}
1433+674 & 49 & 1.76  & 5~10^{-6}    & 1.7 10^2  & 1.83~10^ {-4}  \\ \noalign{\smallskip}
2236-176 & 44 & 1.58  & 5.67~10^{-6} & 1.5 10^2  & 2.0~10^ {-4}   \\ \noalign{\smallskip}
3C465    & 28 & 1.0   & 8.9~10^{-6}  & 99        & 3.2~10^ {-4}   \\ \noalign{\smallskip}
1610-608 & 13 & 0.46  & 1.8~10^{-5}  & 46        & 6.9~10^ {-4}   \\ \noalign{\smallskip}
\noalign{\smallskip}
 \hline
 \hline
 \end{array}
 \]
 \end {table}

\subsection{Turbulent jets}

\label{turbjet} The theory of turbulent jets emerging from a
circular hole can be found in different books with different
theories , see \citet{foot}, \citet{landau}, and \citet{goldstein}.
The basic assumptions  common to the three
already cited approaches  are
\begin{enumerate}
\item  The rate of momentum flow , $J$,
represented by
\begin{equation}
J = constant \times \rho b^2 \overline {v}_{x,max}^2
\quad ,
\end{equation}
is constant; here $x$ is the distance from the initial circular hole
, $b(x)$ is the jet's diameter at  distance $x$,
$\overline{v}_{x,max}$  is  the maximum velocity along the the
centerline, $constant$  is 
\begin{equation}
constant = 2 \pi  \int_0^{\infty} f^2 \xi d\xi 
\quad ,
\label{constant}
\end{equation}
where 
\begin{equation}
f(\xi) =  \frac {\overline {v}_x  }{\overline {v}_{x,max}  } 
 \quad with \quad \xi=\frac{x}{b_{1/2}}
\quad ,
\end{equation}
 and
$\rho$  is the density of the surrounding medium   see equation
(5.6-3) in \citet{foot}. \item The jet's density $\rho$ is constant
over the expansion
      and equal to that  of the surrounding medium.
\end{enumerate}

\begin{figure}
  \begin{center}
\includegraphics[width=10cm]{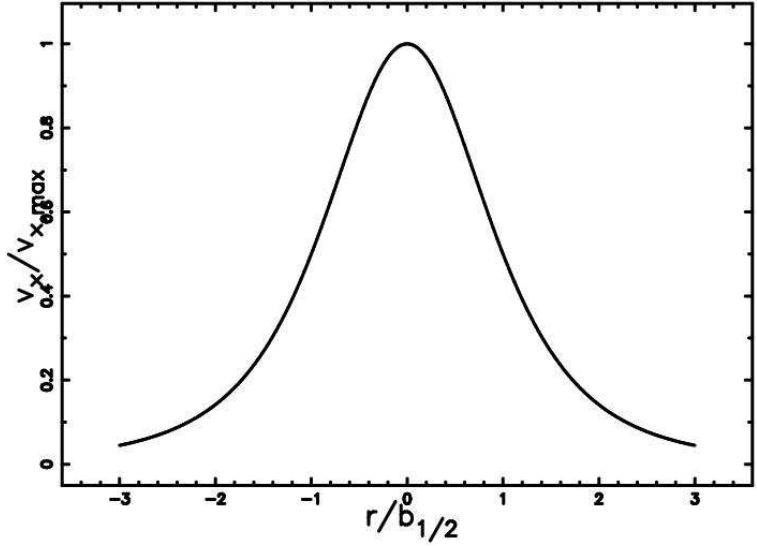}
  \end {center}
\caption {
Mean velocity profile vs. channel radius  
in a circular jet in turbulent flow.
The velocity distribution is a function growing from zero 
(at the wall of the channel) 
to a maximum value in the central region.
          }%
    \label{turbvelocity}
    \end{figure}

When omitting the  details of the computation,  an expression can be found
for the average velocity  $\overline{v}_{x}$,
see equation (5.6-21) in \citet{foot},
\begin{equation}
\overline{v}_{x}  =
\frac{\nu^{(t)} }{x}
\frac {2 C_3^2}{ \bigl [ 1 +\frac {1}{4}(C_3 \frac {r}{x})^2 \bigr ]^2}
\quad ,
\end{equation}
where  $\nu^{(t)}$ is the kinematical
eddy viscosity and  $C_3$ , see equation (5.6-23) in \citet{foot},
\begin{equation}
C_3 =
{\sqrt { \frac {3}{16 \pi}}
{\sqrt { \frac {J} {\rho}}}}
{\frac {1} { \nu^{(t)}}}
\quad .
\label{astroviscosity}
\end{equation}
An important quantity is the radial position , $r=b_{1/2}$ ,
corresponding to an axial velocity one-half of the centerline
value , see equation (5.6-24) in \citet{foot},
\begin{equation}
\frac
{\overline {v}_x (b_{1/2},x) }
{\overline {v}_{x,max}  (x) }
= \frac{1}{2} =
\frac {1}{ \bigl [ 1 +\frac {1}{4}(C_3 \frac {b_{1/2}}{x})^2 \bigr ]^2}
\label{ratiovel}
\quad .
\end{equation}
The experiments in the range of Reynolds number ,$Re$,
$10^4 \le Re \le 3 10^6 $ (see  
\citet{Reichardt1942} ,
\citet{Reichardt1951} and \citet{Schlichting1979})
indicate that
\begin{equation}
b_{1/2}= 0.0848 x
\label{profile}
\quad ,
\label{bvalue}
\end{equation}
and as  a consequence
\begin{equation}
C_3 = 15.17
\label{cm1}
\quad ,
\label{cmvalue1}
\end{equation}
and therefore
\begin{equation}
\frac
{\overline {v}_x (r) }
{\overline {v}_{x,max}  (r) }  =
\frac {1}{ \bigl [ 1 + 0.414 (\frac {r}{b_{1/2}})^2 \bigr ]^2}
\label{ratiovelnumeri}
\quad .
\end{equation}
Figure~\ref{turbvelocity} reports the behaviour of the velocity
distribution as given by equation~(\ref{ratiovelnumeri}).

The average velocity, $\overline{v}_{x}$,
 is   $\approx ~ 1/100$ of the centerline
value when $r/b_{1/2}$ = 4.6  and this allows to say that the
diameter  of the jet is
\begin{equation}
b =  2 \times 4.6 b_{1/2}
\quad .
\end{equation}
On  introducing the opening angle $\alpha$ , the following new 
relationship is found
\begin{equation}
\frac { \alpha }{2} = \arctan \frac { 4.6 b_{1/2}}{x} \quad .
\label{newalfa}
\end{equation}
The generally accepted relationship 
between  the opening angle and Mach number , see equation (A33) in
\citet{deyoung} , is 
\begin{equation}
\frac { \alpha }{2} = \frac{c_s}{v_j} = \frac {1}{M}  \quad ,
\label{oldalfa}
\end{equation}
where $c_s$ is the sound velocity, $v_j$ the jet's velocity 
and  $M$ the Mach number.
The new relationship~(\ref{newalfa})  replaces 
the traditional relationship~(\ref{oldalfa}). 
The parameter $b_{1/2}$  can therefore be connected
with the jet's geometry 
\begin{equation}
b_{1/2}=
\frac {1}{4.6}
\tan (\frac { \alpha }{2}) x
\quad .
\end{equation}
If  this approximate theory  is accepted,
 equation~(\ref{bvalue}) gives
$\alpha=42.61^{\circ}$ : this is the theoretical 
value that originates the so called
Reichardt profiles.
The value of  $b_{1/2}$ fixes the value of $C_3$ and therefore
the eddy viscosity is
\begin{equation}
\nu^{(t)} =
{\sqrt { \frac {3}{16 \pi}}
{\sqrt { \frac {J} {\rho}}}}
{\frac {1} { C_3 }}
=
{\sqrt { \frac {3}{16 \pi}}}
{\sqrt {constant}}
\;
{ b {v}_{x,max}}
{\frac {1}{C_3}}
\quad .
\end{equation}
In order to continue  the integral 
 that appears  in $constant$
 should be evaluated,
 see equation~(\ref{constant}).
The numerical integration gives 
\begin{equation}
\int_0^{\infty} f^2 \xi d\xi =0.402 
\quad ,
\end{equation}
and therefore
\begin{equation}
constant=2.528 
\quad .
\end{equation}
On introducing typical parameters of jets  like
$\alpha$=$5^{\circ}$,
${v}_{x,max}$=$v_{100}=v[{\mathrm{km/sec}}]/100$, $b=b_1$ where $b_1$ is the
momentary radius in ${\mathrm{ kpc}}$ ,
it 
is possible to deduce an astrophysical  formula for
the kinematical eddy viscosity
\begin{equation}
\nu^{(t)} =
0.0028\, b_1 v_{100}  \mathrm{ \frac {kpc^2}{10^7 year}}
\quad  when
\quad C_3 = 135.61
\quad .
\end{equation}

This paragraph  concludes underlining   the fact   that
in extragalactic sources it is possible to observe 
both a small opening angle  , $\approx~5^{\circ}$, see
Section~\ref{secwat} and 
great opening angles , i.e. $\approx~34^{\circ}$ in the outer 
regions of 3C31 \citet{LaingBridle2004}.

\section{A continuous  model }

\label{sezione_a}
\label {sezione_cont}
The jet is now explained
as due to  a continuous flow in a given
direction.

The constrained basic elements of our theory are  the mean spread
rate of the jet measured on the radio  maps , the linear density
of energy and  the initial  jet radius $r_i$. The turbulent
hyper-sonic  flow , Mach number $>$ 6 , is one of the most
complicated problems in fluid mechanics. Here the key assumption
is that due to the turbulent mass transfer, the density in the jet
is the same as  the surrounding medium, see for example equation
(10.25) in ~\citet{hughes}.
This allows  the jet law of motion to be expressed 
in terms of the IGM's
density.
We identify  our jet with a pyramidal section
characterised by a solid   angle   $\Delta\;\Omega$
and overall length $x_1$.
From a practical point of view,
 $\alpha_0$ (the first opening angle)
is reported in the captions.
The solid angle $d\Omega$ in spherical  coordinates
(r,$\theta , \phi $) is
\begin{equation}
d \Omega = \sin (\theta) d\theta d \phi \quad .
\end {equation}
On shifting
to the finite differences, the following is obtained
\begin{equation}
\Delta \Omega
 =
(\cos (0) - \cos (\alpha_0 \frac{2 \pi}{360}  ) ) (\alpha_0 \frac{2
\pi}{360}) ~~         \mathrm {steradians} 
\label{solid}
\quad ,
\end {equation}
that when the angles are small becomes
\begin{equation}
\Delta \Omega
 \approx
\frac{1}{2} (\frac {\alpha_0 2 \pi }{360^{\circ }})^3 \mathrm{steradians} 
\label{solidsmall}
\quad ,
\end {equation}
where  $\alpha_0$  represents the limit of the integration
and  is expressed in degrees.

\begin{figure}
  \begin{center}
\includegraphics[width=10cm]{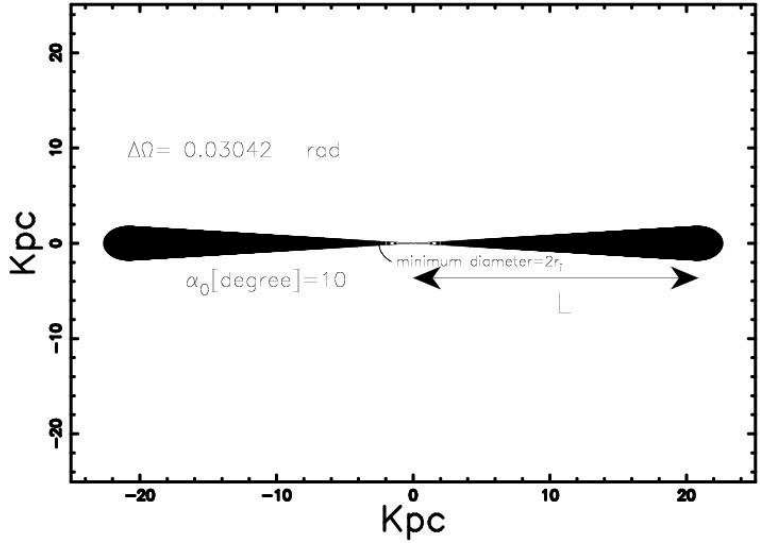}
  \end {center}
\caption {
Two dimensional view of the continuous
jet; the physical parameters
are the same as  the simulation of NGC4061 ,
see Figure~\ref{NGC4061}.
          }%
    \label{jet_cont}
    \end{figure}

As an example,  when   $\alpha_0$=$10^{\circ}$,
 $\Delta\;\Omega=2.65~10^{-3}~\mathrm{steradians}$.
A plot  that depicts
the assumed geometry is reported in Figure~\ref{jet_cont}.

The approximate advancing area of the jet of
momentary length $x$,
$A_{j}$,  will therefore be
\begin{equation}
A_{j} \approx
x^2 \Delta \Omega
\quad  .
\end {equation}
The jet  is divided in many
slices perpendicular  to the direction
of motion , each one of thickness $r_i$~(the initial radius);
the  number of slices, $n_j$, is therefore
\begin{equation}
n_j  = NINT (\frac{x_1}{r_i})
\quad ,
\end {equation}
where NINT denotes the nearest integer.
The volume
of  slice,  $V_s$,  is
\begin{equation}
V_{s} = x^2  \Delta \Omega  r_i
\quad ,
\label{eq_volume_s}
\quad
\end{equation}
and the  mass acquired in a slice  from the external medium
of a density, $\rho$, is
\begin{equation}
\Delta M =  \rho \times V_s = \rho   x^2  \Delta \Omega  r_i
\quad .
\end{equation}
The hypothesis of a jet in which  $\rho$ is  constant, 
is used in the
theory of turbulent jets, see Section~\ref{turbjet}.
 The  equation that represents
 the momentum conservation is:
\begin {equation}
\frac{d}{dt}\left(\Delta M \dot{x}\right)=p x^2
\Delta \Omega
\quad ,
\label{eq_momentum_c}
\end {equation}
where $p$  is the pressure. 
In the case of constant density,
equation~(\ref{eq_momentum_c}) becomes
\begin {equation}
\frac{d}{dt}\left(x^2 \dot{x}\right)=
\frac {p x^2 }{\rho r_i}
\quad .
\label{eq_diff}
\end {equation}

The pressure of  internal gas decreases according to
the adiabatic law
\begin {equation}
p=p_i({V_i \over V})^{5/3}
\label {eq:adiabatic1}
\quad ,
\end {equation}
where  $V_i$ is  the volume of the first slice and
\begin {equation}
p_i =  \frac {2}{3}  \frac {\frac {E}{n_j} }
    {  V_i}
\quad .
\label {eq:adiabatic3}
\end {equation}

Under the adiabatic hypothesis,  the differential
equation~(\ref{eq_diff}) is transformed  in
\begin {equation}
\frac{d}{dt}\left(x^2 \dot{x}\right)= H x^{- 4/3} \quad ,
\label{eq_diff_d}
\end {equation}
where
\begin{equation}
H = \frac {1}{\rho} p_i \frac  {1} { r_i^{-7/3}
\Delta\Omega^{5/3}}
\quad ,
\end {equation}
and
\begin{equation}
p_i = \frac{\frac{2}{3} E} {x_1 r_i^2}
\quad .
\end {equation}

To integrate this equation, $x^2 \dot{x}=Ax^{\alpha}$ is used.
After adopting the initial condition of  $x=0$ at $t=0$
and assuming $\rho$ is constant,
the equation
representing
 the third law of motion  is obtained
\begin {equation}
x=\left( \frac{13}{18} \frac{(3 \alpha +4)}{\alpha} H t^2 \right )
^{3/13} \quad , \label{third}
\end{equation}
and  since $\alpha=5/6$ :
\begin {equation}
x=\left( \frac{507}{90} H t^2 \right ) ^{3/13} \quad .
\end{equation}
The velocity  of the jet  turns out to be:
\begin {equation}
v(t)= {\dot x} =  \frac  {6}{13} \frac {x} {t}
\quad.
\label{eq:velocity_ada}
\end {equation}
The  astrophysical quantities can now be introduced
\begin {equation}
x(t)=2.541\,{\mathrm{kpc}}\, \left( {\frac {{{\it t_7}}^{2}{\it E_{56}}\,
\sqrt [3]{{\it r_i^1}}}{{\it x_1^1}\,{\Delta\Omega}^{5/3}{\it n_0}}}
\right) ^{3/13}
\label{eq:astror}
\quad ,
\end {equation}
where  $E_{56}$ =$E/10^{56}$ \mbox {ergs} ,
$r_i^1= r_i/1{\mathrm{kpc}}$ , $x_1^1=x_1/1{\mathrm{kpc}}$ ,
$t_7$ = $t/(10^7)$ \mbox {year} and
$\rho$ = $n_0~m_H$
with   $n_0$  representing  the number of particles in a
cubic centimetre and $m_H$ the
hydrogen  mass.
A similar formula is deduced  for the velocity
\begin {equation}
\dot R =114.92 \,{\mathrm{\frac{km}{s}}}\,
\left( {\frac {{\it E_{56}}\,
\sqrt [3]{{\it r_i^1}}}{{\it x_1^1}\,{\Delta\Omega}^{5/3}{\it n_0}}}
\right) ^{3/13} t_7 ^{-7/13}
\label{eq:astrov}
\quad .
\end {equation}
In the case of SNR  a comparison can  be made with the radius
$R(t)\propto~t^{2/7}$ of the adiabatic  phase of the
SNR , see  for example
~\citet{Dalgarno1987}  , and with  $R(t)\propto~t^{2/5}$ 
of the so called Sedov solution , see \citet{Sedov1959} and
\citet{landau}:
\begin{equation}
R(t)=
\bigl ({\frac {25}{4}}\,{\frac {{\it E}\,{t}^{2}}{\pi \,\rho}}\bigr)^{1/5}
\label{rsedov}
\quad ,
\end{equation}
where  $t$ represents the time, 
$E$ is the energy injected in the explosion,
and $\rho$ is the density of matter.
When the jets are analyzed  a comparison can be made  
with the  solution  presented in
\citet{Kaiser1997}.
In their model  the   density $\rho_x$  
 of the gas surrounding the jets  
 scales as
\begin{equation}
\rho_x = \rho_0 (\frac {d}{a_0}) ^{-\beta} 
\quad  ,
\end{equation}
where $d$ is the radial distance from the core of the source ,
$\rho_0$ is the initial  density ,
$a_0$ is the scale length  and  $\beta$ a parameter
comprised between 0 and 2.
Their 
 length of the jet $L_j$  scales  
 as
\begin{equation}
L_j \propto t^{3/5 -\beta} 
\quad ,
\end {equation}
and  explains the division 
between FRI and FRII objects in jet power.

\section{Two--phase continuous model }

\label{twophase}

The starting point is the conservation of the
momentum's flux in a "turbulent  jet"  as outlined
in  \citet{landau} (pag.~147).
The initial  point   is  characterised  by the following
section  :
\begin {equation}
A_0=\pi~r_0^2
\quad  .
\end{equation}
Once $\alpha_0$ the first opening angle ,
$x_0$ the initial position on the $x$--axis and
$v_0$ the initial velocity are introduced,
 section  $A$ at  position $x$  is
\begin {equation}
A(x)=\pi \bigl ({\it r_0}+ \left( x-{\it x_0} \right) \tan \left( (1/2)\,\alpha_{{0}}
 \right) \bigr )^2
\quad  .
\end{equation}
The conservation  of the total momentum flux states that
\begin{equation}
\rho  v_0^2 A_0 =
\rho  v(x)^2 A(x)
\quad  ,
\label{conservazione}
\end {equation}
where $v(x)$ is the velocity at  position $x$ .
The previous  equation  is valid if
the pressure along
the jet is constant and viscosity effects 
are either not considered or have 
the same effect in the
whole momentum flux along the jet.  
Due to the turbulent transfer, the density $\rho$
is the same on both the two sides of
equation~(\ref{conservazione}).
The trajectory of the jet  as a function of the time
is easily  deduced
from equation~(\ref{conservazione})
\begin {eqnarray}
x= -{\frac {-{\it x_0}\,\tan \left( (1/2)\,\alpha_{{0}} \right) +{\it r_0}-
\sqrt {{\it r_0}\, \left( {\it r_0}+2\,\tan \left( (1/2)\,\alpha_{{0}}
 \right) {\it v_0}\,t \right) }}{\tan \left( (1/2)\,\alpha_{{0}} \right)
}} \quad  ,
\label{traiettoria}
\end{eqnarray}
and this can be considered the fourth law of motion.

The  velocity turns out to be
\begin{equation}
{\it v(t)}={\frac {{\it v_0}\,{\it r_0}}{\sqrt {{\it r_0}\, \left( {\it r_0}
+2\,\tan \left( (1/2)\,\alpha_{{0}} \right) {\it v_0}\,t \right) }}}
\quad  .
\label{velocita}
\end {equation}
In   the applications  of equation~(\ref{traiettoria}) and
(\ref{velocita})
  $r_0$ , $r$  ,  $x_0$ and $x$
will be expressed in $\mathrm{kpc}$ , the time $t$ in units of
$10^7~{\mathrm{year}}$ ,  $v_0$ and $v(x)$ in
$\mathrm {\frac{kpc}{10^7~year}}$.
The  velocity  $v_0$ can be parametrised as a function
of the light velocity c as  $v_0=\frac{c}{c_f}$
where $c_f$  is an input parameter allowed to vary
between 1.2 and 10.

It is now possible to introduce the two phase beam in which  , due
to a certain physical  phenomena, for example
the evolution of the K-H  instabilities , the beam
abruptly increases  the opening angle that from
$\alpha_0$  (first opening angle) becomes  $\alpha_1$
 (second opening angle).
This phenomena  happens
at a given time $t_1$ ( $t_1~<~t_{2}$ ) and
a corresponding length $x_1$ (  $x_1~<~x_2$):
here $t_2$ denotes the age of the radio-source
and  $x_2$ its   global length.
In this  second  phase  the
trajectory  is
\begin {equation}
x= -{\frac {-{\it x_1}\,\tan \left( (1/2)\,\alpha_{{1}} \right) +{\it r1}-
\sqrt {{\it r1}\, \left( {\it r1}+2\,\tan \left( (1/2)\,\alpha_{{1}}
 \right) {\it v_1}\,(t-t_1) \right) }}{\tan \left( (1/2)\,\alpha_{{1}} \right)
}}
\quad  ,
\label{traiettoria_2}
\end{equation}
and the velocity
\begin{equation}
{\it v(t)}={\frac {{\it v_1}\,{\it r_1}}{\sqrt {{\it r_1}\, \left( {\it r_1}
+2\,\tan \left( (1/2)\,\alpha_{{1}} \right) {\it v_1}\,(t-t_1) \right) }}}
\quad .
\label{velocita_2}
\end {equation}

\section{Two--phase relativistic  model }

\label{twophaserelativistic}
The starting point is the energy momentum tensor  ,$ T^{ik} $,
\begin{equation}
T^{ik} = w u^i u^k - p g^{ik}
\quad  , 
\end{equation}
where $u^i$ is the 4-velocity ,
$i$ varies  from 0 to 3 ,
$w$ is the enthalpy for unit volume ,
$p$ is the pressure and 
$g^{ik}$ the inverse metric of the manifold, see \citet{landau}.
The condition for momentum conservation 
in the presence of velocity , $v$, along one direction  
states that
\begin{equation}
(w (\frac{v}{c})^2 \frac { 1}{ 1 -\frac {v^2}{c^2} } +p) A = cost
\quad ,
\label{enthalpy}
\end{equation}
where $A$ is the considered area in the direction perpendicular
to the motion. The enthalpy for unit volume is  
\begin{equation}
w= c^2 \rho  + p 
\quad ,
\end{equation}
where $\rho$ is the   density ,
and $c$ the light velocity.
The reader  may be puzzled by the
$\Gamma^2$ factor in equation~(\ref{enthalpy}),
where $\Gamma^2= \frac { 1}{ 1 -\frac {v^2}{c^2} }$.
 However it should be remembered  that
$w$ is not an enthalpy, but an enthalpy per unit volume: 
the extra $\Gamma$
factor arises from  "length contraction" in the direction of motion.

With the assumption of turbulent jets ,$p=0$ ,
the momentum conservation law
is obtained 
\begin{equation}
(\rho  v^2 \frac { 1}{ 1 -\frac {v^2}{c^2} }) A = cost
\quad .
\end{equation}
This equation is equal to equation~(A.32) , the jet "thrust",   
in ~\citet{deyoung}.
Note the similarity between the previous formula and 
condition (135.2) in \citet{landau} concerning 
the shock waves : when $A$=1 they are equals.

In two sections of the jet we have :
\begin{equation}
\rho\,{{\it v_0}}^{2}\pi \,{{\it r_0}}^{2}{\frac {1}{ 1-{\frac {{{
\it v_0}}^{2}}{{c}^{2}}}}}=\rho\,{v}^{2} \left( (\pi \,{{\it r_0}}^{2}+
\pi \,{\it r_0}\,x\alpha_0+O \left( {\alpha_0}^{2} \right) ) \right) {
\frac {1}{ 1-{\frac {{v}^{2}}{{c}^{2}}}}}
\quad ,
\label{conservazionerel}
\end {equation}
where $v$ is the velocity at  position $x$, $v_0$ the velocity 
at $x$=0 , $c$ the light velocity  and $\alpha_{{0}}$ 
the opening angle  of the jet.
Also here due to the turbulent transfer, the density $\rho$
is the same on both  sides of
equation~(\ref{conservazionerel}) and the  following
second degree equation in $\beta=\frac{v}{c}$ is obtained:
\begin{eqnarray}
{{\it \beta}}^{2}{\it r_0}+{{\it \beta}}^{2}x\alpha_0-{{\it \beta}}^{2}{{
\it \beta_0}}^{2}x\alpha_0-{{\it \beta_0}}^{2}{\it r_0}=0
\end{eqnarray}
where $\beta_0= \frac{v_0}{c}$.
The positive solution is :
\begin{equation}
\beta=   
{\frac {\sqrt {{\it r_0}}{\it \beta_0}}{\sqrt {{\it r_0}+x\alpha_0-{{\it 
\beta_0}}^{2}x\alpha_0}}}
\quad .
\label{beta}
\end{equation}
From equation~(\ref{beta}) it is possible to deduce the distance 
$x_{1/2}$ after 
which the velocity is half of the  initial value:
\begin{equation}
 x_{1/2} =  
-3\,{\frac {{\it r_0}}{\alpha_0\, \left( -1+{{\it \beta_0}}^{2} \right) }}
\quad . 
\label{betamezzi}
\end {equation}
In FR-II's  radio-galaxies  the
velocity is thought to be  relativistic until termination at hot-spots;
formula~(\ref{betamezzi})  states that the typical distance over
which the jet is relativistic is a  function of the three parameters
$\alpha_0$ , $r_0$  and  $\beta_0$.

The trajectory of the relativistic jet  as a function of the time  can  be
  deduced
from equation~(~\ref{beta}) and is  
\begin {equation}
\int _{0}^{x}
\frac{ 1}{{\frac {\sqrt {{\it r_0}}{\it \beta_0}}{\sqrt {{\it r_0}+x\alpha_0-{{\it 
\beta_0}}^{2}x\alpha_0}}}
} dx =ct
\quad  .
\end{equation}
On integrating it is possible to  obtain the equation of the trajectory
\begin{equation}
2/3\,{\frac {{{\it r_0}}^{3/2}- \left( {\it r_0}+x\alpha_0-{{\it \beta_0}}^{
2}x\alpha_0 \right) ^{3/2}}{\alpha_0\, \left( -1+{{\it \beta_0}}^{2}
 \right) \sqrt {{\it r_0}}{\it \beta_0}}}-ct=0
\quad . 
\label{firstreltraj}
\end{equation}
After some manipulation equation~(\ref{firstreltraj})
becomes 
\begin{eqnarray}
 \left( -3\,{\alpha_0}^{3}{{\it \beta_0}}^{2}+3\,{\alpha_0}^{3}{{\it \beta_0}}
^{4}-{{\it \beta_0}}^{6}{\alpha_0}^{3}+{\alpha_0}^{3} \right) {x}^{3}+
  \nonumber\\
 \left( 3\,{\it r_0}\,{\alpha_0}^{2}-6\,{\it r_0}\,{\alpha_0}^{2}{{\it \beta_0
}}^{2}+3\,{\it r_0}\,{{\it \beta_0}}^{4}{\alpha_0}^{2} \right) {x}^{2}+
 \nonumber\\
 \left( -3\,{{\it r_0}}^{2}{{\it \beta_0}}^{2}\alpha_0+3\,{{\it r_0}}^{2}
\alpha_0 \right) x   -                         
 \nonumber\\
9/4\,{c}^{2}{t}^{2}{\alpha_0}^{2}{\it r_0}\,{{\it \beta_0}
}^{6}-3\,{{\it r_0}}^{2}ct\alpha_0\,{\it \beta_0}+3\,{{\it r_0}}^{2}ct\alpha_0
\,{{\it \beta_0}}^{3}-      \nonumber\\
9/4\,{c}^{2}{t}^{2}{\alpha_0}^{2}{\it r_0}\,{{\it 
\beta_0}}^{2}+9/2\,{c}^{2}{t}^{2}{\alpha_0}^{2}{\it r_0}\,{{\it \beta_0}}^{4}
=0 
\quad .   
\label{reltraj}
\end{eqnarray}
The parameter $p$ , see Appendix~\ref{appendix_cubic}, 
that  regulates the solutions of the cubic equation
is now evaluated.
When the equation~(\ref{reltraj}) is considered 
we have  $p=0$ and therefore we have one real  root
that  is (the  fifth law  of motion ) :
\begin {equation}
x(t)= 
-1/2\,{\frac {-2\,{\it r_0}+\sqrt [3]{2}\sqrt [3]{{\it r_0}} \left( 2\,{
\it r_0}+3\,ct\alpha_0\,{\it \beta_0}-3\,ct\alpha_0\,{{\it \beta_0}}^{3}
 \right) ^{2/3}}{\alpha_0\, \left( -1+{{\it \beta_0}}^{2} \right) }}
\label{traiettoriarel}
\quad .
\end{equation}
The velocity of the relativistic jet  as function 
of the time is 
\begin{equation}
v(t) = 
{\frac {c{\it \beta_0}\,\sqrt [3]{{\it r_0}}\sqrt [3]{2}}{\sqrt [3]{2\,{
\it r_0}+3\,ct\alpha_0\,{\it \beta_0}-3\,ct\alpha_0\,{{\it \beta_0}}^{3}}}}
\quad .
\end{equation}
The  velocity  $v_0$ can be parametrised as a function
of the  parameter $\beta_0$ ,  
a  parameter allowed to vary
between 0.1 and  0.99999.
The two  previous equations  can  be expressed 
in  astrophysical units 
once  $r_0^1=r_0/1{\mathrm{kpc}}$ and 
$t_7$ = $t/(10^7\mathrm{year})$   are introduced 
\begin{equation}
x(t) =  
1/2\,{\frac {2\,{\it r_0^1}-\sqrt [3]{2}\sqrt [3]{{\it r_0^1}} \left( 2\,{
\it r_0^1}+ 9180.0\,{\it t_7}\,\alpha\,{\it \beta_0}- 9180.0\,{\it t_7}\,
\alpha\,{{\it \beta_0}}^{3} \right) ^{2/3}}{\alpha\, \left( -1+{{\it 
\beta_0}}^{2} \right) }}{\mathrm{ kpc}}
\quad ,
\end{equation}
and 
\begin{equation}
v(t)=
299788.2\,{\frac {{\it \beta_0}\,\sqrt [3]{{\it r_0^1}}\sqrt [3]{2}}{\sqrt
[3]{2\,{\it r_0^1}+ 9180.0\,{\it t_7}\,\alpha\,{\it \beta_0}- 9180.0\,
{\it t_7}\,\alpha\,{{\it \beta_0}}^{3}}}} {\mathrm{\frac{km}{s}}} 
\quad .
\end{equation}
The transition from relativistic to classical velocity 
is still represented by formulas~(\ref{traiettoria_2}) and
(\ref{velocita_2})  but $x_1$,$r_1$ and $v_1$ are the parameters at the
end of the relativistic phase.

\section{The change of framework  }
\label{sezione_c}
The wide spectrum of observed morphologies that characterises the
head--tail radio  galaxies could be due to the  kinematical
effects as given by  the composition of the velocities in
different kinematical frameworks such as decreasing jet velocity ,
jet precession, rotation, and proper  velocity of the host galaxy
in the cluster. These effects were partially  analysed in
~\citet{zan1} ; here   part of the developed  theory and symbols
will be used, and will now be briefly   reviewed in order to
represent the law of  motion through a matrix. Of particular
interest  is the  evaluation of  various matrices that  will
enable us to cause  transformation from  the inertial coordinate
system of the jet to the coordinate system in which the host
galaxy is  moving in  space. 
The various coordinate systems will be
${\bf x}$=$(x,y,z$) , ${\bf x}^{(1)}$=$(x^{(1)},y^{(1)},z^{(1)})$
, $\ldots$ ${\bf x}^{(5)}$=$(x^{(5)},y^{(5)},z^{(5)})$. The
vector representing the  motion of the jet  will be represented by
the following $1 \times 3$  matrix
\begin{equation}
G=
 \left[ \begin {array}{c} 0\\\noalign{\medskip}0\\\noalign{\medskip}{
\it L(t)}\end {array} \right] 
\quad  ,
\end{equation}
where the jet motion L(t) is considered along the z-axis.

The  jet axis, $z$, is inclined at an angle $\Psi_{prec}$
relative to an axis $z^{(1)}$ and therefore
the  $3 \times 3$  matrix,
representing    a rotation through the x axis,
is given by:
\begin {equation}
F=
 \left[ \begin {array}{ccc} 1&0&0\\\noalign{\medskip}0&\cos \left( 
\Psi_{prec} \right) &\sin \left( \Psi_{prec} \right) \\\noalign{\medskip}0&-\sin
 \left( \Psi_{prec} \right) &\cos \left( \Psi_{prec} \right) \end {array}
 \right] 
\quad  .
\end {equation}
From a practical point of view $\Psi_{prec}$ can be derived  by
measuring the half opening angle of the maximum of the sinusoidal
oscillations (named wiggles) that characterises the jet.

If the jet is undergoing precession around the
$z^{(1)}$ axis,  $\Omega_{prec}$ can be
the angular velocity of precession expressed in
$\mathrm{radians}$ per unit time ; 
 $\Omega_{prec}$ is computed from the radio  maps
by measuring the
number of sinusoidal oscillations
that characterise  the jet.
The  transformation from the coordinates
${\bf x}^{(1)}$ fixed in the frame of the
precessing jet to the non-precessing coordinate
${\bf x}^{(2)}$
is represented by the  $3 \times 3$  matrix
\begin{equation}
  E=
 \left[ \begin {array}{ccc} \cos \left( \Omega_{prec}\,t \right) 
&\sin \left( \Omega_{prec}\,t \right) &0\\\noalign{\medskip}-\sin
 \left( \Omega_{prec}\,t \right) &\cos \left( \Omega_{prec}
\,t \right) &0\\\noalign{\medskip}0&0&1\end {array} \right] 
\quad  .
\end{equation}
The arbitrary orientation of the precession axis relative to the
central galaxy at rest,  necessitates a transformation from the
coordinate frame ${\bf x}^{(2)}$ to the frame   ${\bf x}^{(3)}$.
The relative orientations are assumed to be characterised by the
Euler  angles $(\Phi_j, \Theta_j, \Psi_j)$.
There is not uniform agreement on the designation of the Euler angles and
the manner in which they are generated. We have chosen here to use the
conventions found  in ~\citet{Goldstein2002} 
and the $3 \times 3$  matrix ,
$D_E$,   is 
\begin  {eqnarray}
\lefteqn {D_E= }\nonumber  \\
&  \left [ \matrix {cos \Psi_j cos \Phi_j  - cos\Theta_j sin \Phi_j sin\Psi_j
      & cos \Psi_j sin \Phi_j  + cos\Theta_j cos \Phi_j sin\Psi_j
      &sin\Psi_jsin\Theta_j  \cr
      -sin \Psi_j cos \Phi_j  - cos\Theta_j sin \Phi_j cos\Psi_j
      & -sin \Psi_j sin \Phi_j  + cos\Theta_j cos \Phi_j cos\Psi_j
      &cos\Psi_jsin\Theta_j \cr
      sin\Theta_j sin\Phi_j &-sin\Theta_j cos\Phi_j 
      &cos \Theta_j    \cr
}
 \right ] 
\quad  .
\end {eqnarray}
On assuming  that $
\Phi_{j} $=  $0^{\circ}$ , $ \Psi_{j} $=  $0^{\circ}$ and  $
\Theta_{j} $= $90^{\circ}$~
we have the simple expression 
\begin{equation}
  D_E =
 \left[ \begin {array}{ccc} 1&0&0\\\noalign{\medskip}0&0&1
\\\noalign{\medskip}0&-1&0\end {array} \right] 
\quad  .
\end{equation}

Another $3 \times 3$  matrix  is introduced
which  represents the transformation from
 $(\bf{x}^3)$ to  $(\bf{x}^4)$ and
defined  by a rotation through the axis $z^{(3)}$ of an angle
$\Omega_G t $ where $\Omega_G $ is the angular
velocity of the
galaxy expressed
in $\mathrm{radians}$/time units; the total angle of rotation of the galaxy
being  $\alpha $ =
$\Omega_G\; t_{max}$ :
\begin{equation}
  C =
 \left[ \begin {array}{ccc} \cos \left( \Omega_{G}\,t
 \right) &\sin \left( \Omega_{G}\,t \right) &0
\\\noalign{\medskip}-\sin \left( \Omega_{G}\,t \right) &\cos
 \left( \Omega_{G}\,t \right) &0\\\noalign{\medskip}0&0&1
\end {array} \right] 
\quad  .
\end{equation}
In the astrophysical applications
$\Omega_G=0$ because the rotation period of a typical
elliptical  galaxy is less than the lifetime
of a typical radio  source; only in the case of NGC326
$\Omega_G~\neq~0$,  see  discussion in Section~\ref{NGC326sec}.

The last  translation represents
the change of framework from
$\bf(x^{(4)})$, which is co-moving with
the host galaxy, to a system
$\bf (x^{(5)})$ in comparison  to which the
host galaxy is in uniform
motion.
It should be  remembered  that  the dispersion velocity  in the cluster
(not to be confused with the recession velocity) is
$\approx$ 600 ${\mathrm{km/sec}}$ ,
see  Table~1
in
~\citet{venkatesan}
.
The relative motion of the origin of the coordinate system
$(x^{(4)},y^{(4)},z^{(4)})$ is defined by the
Cartesian components of the galactic velocity ${\bf v_G}$,
and  the  required  $1 \times 3 $  
matrix transformation representing  this translation is:
\begin{equation}
  B=
 \left[ \begin {array}{c} v\_x\\\noalign{\medskip} v\_y\\\noalign{\medskip}{
\it v\_z}\,t\end {array} \right] 
\label {transla}
\quad  .
\end{equation}
On assuming, for the sake of simplicity, that $v_y$=0  and $v_x$=0,
the translation matrix  becomes:
\begin{equation}
  B=
 \left[ \begin {array}{c} 0\\\noalign{\medskip}0\\\noalign{\medskip}{
\it v\_z}\,t\end {array} \right] 
\quad  .
\end{equation}
In other words,  the direction of the galaxy motion in the IGM and
the direction of the jet are perpendicular. From a practical point
of view   the galaxy velocity can be measured by dividing the
length of the radio galaxy in a direction perpendicular to the
initial jet velocity by the lifetime of the radio  source; an
example of such measurement is reported  in Section~\ref{galvel}.

 The
final $1 \times 3$  matrix $A$  representing  the
``motion law'' can be found
by composing  the six  matrices  already
described
\begin  {eqnarray}
\lefteqn {A =   B + ( C \cdot D_E \cdot E \cdot F \cdot G)  } \nonumber  \\
 =&
 \left[ \begin {array}{c}  \left( \cos \left( \Omega_{G}\,t
 \right) \sin \left( \Omega_{prec}\,t \right) \sin \left( \Psi_{prec}
 \right) +\sin \left( \Omega_{G}\,t \right) \cos \left( 
\Psi_{prec} \right)  \right) {\it L(t)}\\\noalign{\medskip} \left( -\sin
 \left( \Omega_{G}\,t \right) \sin \left( \Omega_{prec}
\,t \right) \sin \left( \Psi_{prec} \right) +\cos \left( \Omega\,{\it 
galaxy}\,t \right) \cos \left( \Psi_{prec} \right)  \right) {\it L(t)}
\\\noalign{\medskip}{\it v\_z}\,t-\sin \left( \Psi_{prec} \right) \cos
 \left( \Omega_{prec}\,t \right) {\it L(t)}\end {array} \right] 
\quad  .
\end   {eqnarray}
The  three components of the previous
 $1\times 3$  matrix  $A$
represent the jet motion
along the Cartesian coordinates as given by the observer
that sees the galaxy moving  in uniform motion.
\section{Application of the continuous model}

\label{sezione_d} The continuous beam model developed in
Section~\ref{sezione_cont}
 is basically represented
by equations~(\ref{eq:astror}) and  (\ref{eq:astrov})
and
can be easily implemented in a code;
this is  now applied to various radio--sources.

An attempt to reproduce the mechanical power dependence in the jet
trajectory  is carried out in Section~\ref{sectransition}.

\subsection{The first part of  NGC1265}

\begin{figure}
  \begin{center}
\includegraphics[width=10cm]{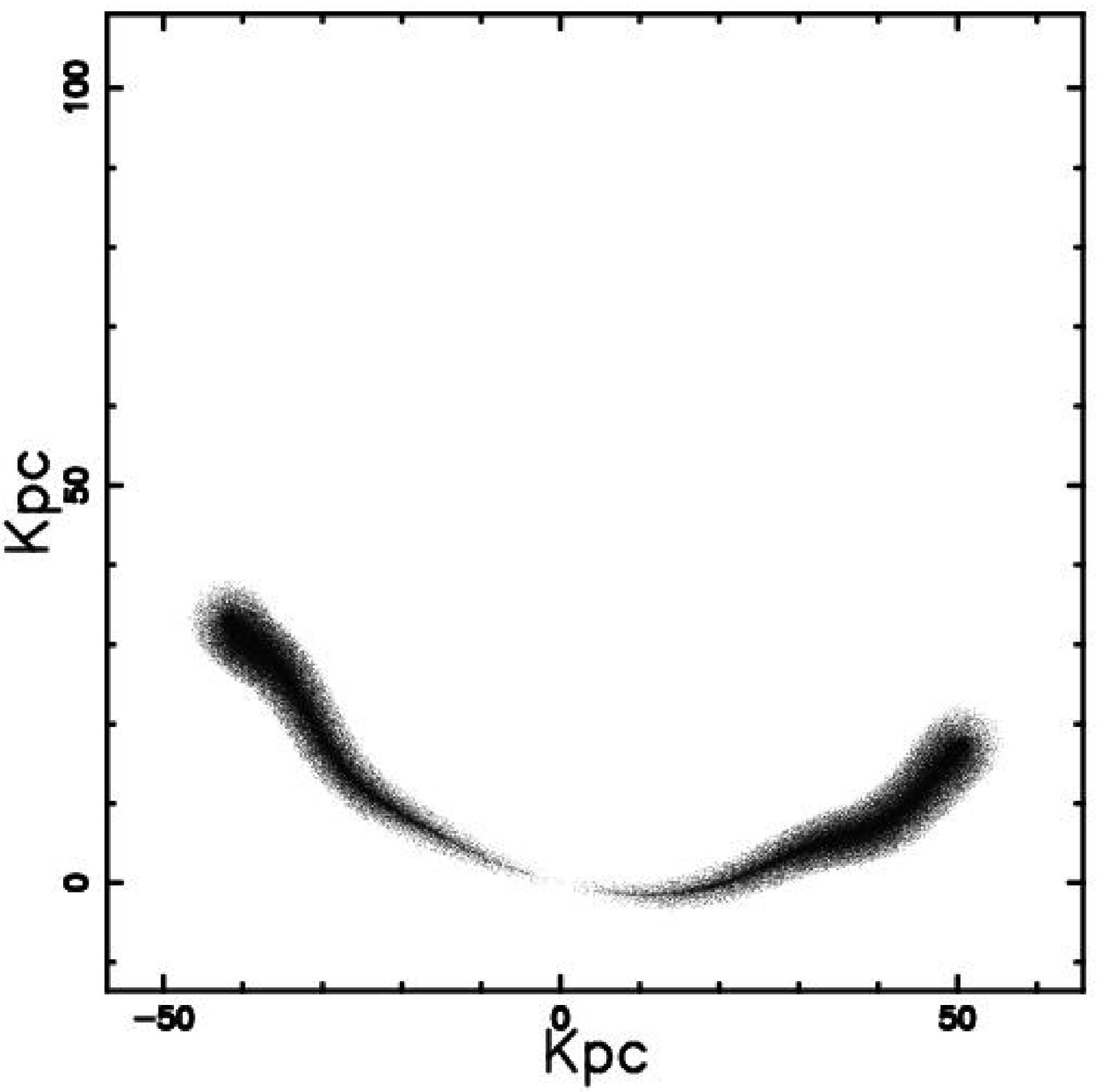}
  \end {center}
\caption {
 Continuous three-dimensional trajectory of NGC1265:
 the three Eulerian angles (English notation)
 characterising the point of view are $ \Phi $= 80   $^{\circ }$
 , $ \Theta $= 80   $^{\circ }$
 and  $ \Psi $= 10   $^{\circ }$.
 The precession is characterised by the angle
  $ \Psi_{prec} $=  2   $^{\circ }$
 and by the angular velocity
 $ \Omega_{prec} $= 54.00 [$^{\circ}/10^7\mathrm{year}$].
 The three Eulerian angles are:
 $ \Phi_{j} $=  0   $^{\circ }$
 , $ \Psi_{j} $=  0   $^{\circ }$
 and  $ \Theta_{j} $= 90   $^{\circ }$.
 The angle of rotation of the galaxy is
 $ \alpha_{G} $=  0   $^{\circ }$.
 The physical parameters characterising the jet motion
 are : $E_{56}$=  0.02,
 $t_1$= 10.00~$10^7\mathrm{year}$,
 $n_0$=  0.01~$\mathrm{particles/cm^3}$,
 $x_1$= 18.00~${\mathrm{kpc}}$ ,
 $r_i$=  0.90~${\mathrm{kpc}}$ ,
 $\alpha_0$= 10   $^{\circ }$.
}
\label{NGC1265_proj}
\end{figure}

\begin{figure}
  \begin{center}
\includegraphics[width=10cm]{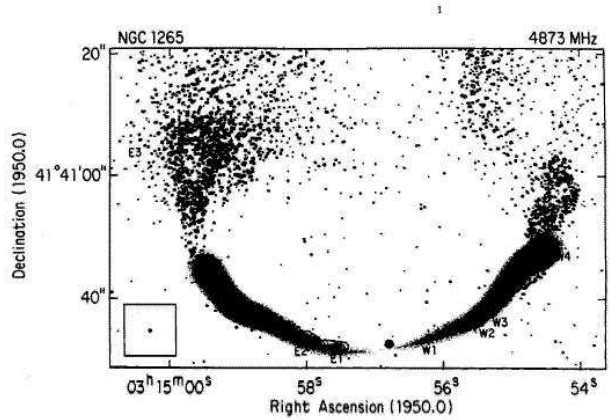}
  \end {center}
\caption {
Figure 1 of ~\citet{odea} representing NGC1265 at 4873 MHz superimposed on 
the theoretical trajectory represented by  Figure~\ref{NGC1265_proj}.
}
\label{NGC1265_proj_theo}
\end{figure}

The first target chosen  for  our  simulation
was  the radio source NGC1265 at 6 cm.
Some basic input parameters
extracted from
~\citet{odea}
 are reported in
Table~\ref{data_ngc1265}.
 \begin{table}
 \caption[]{T\lowercase{he data from the radio  map of }NGC1265  }
 \label{data_ngc1265}
 \[
 \begin{array}{cc}
 \hline
 \hline
 \noalign{\smallskip}
 \mathrm {L_{rad}[ergs/sec]} & mean ~spread~rate [degree] \\
 \noalign{\smallskip}
 \hline
 \noalign{\smallskip}
1.1 10^{40} &  10 \\ \noalign{\smallskip}
\noalign{\smallskip}
 \hline
 \hline
 \end{array}
 \]
 \end {table}
The result of the simulation is reported
in Figure~\ref{NGC1265_proj}  and
in Figure~\ref{NGC1265_proj_theo} with the radio map 
superimposed on the simulation ;
the  adopted input parameters
 are given  in the caption.

The output data as obtained from the code concerning
 the overall
jet length,
the perpendicular-displacement, and  the galaxy
velocity  of NGC1265 are reported in Table~\ref{fine_ngc1265}.
 \begin{table}
 \caption[]{T\lowercase{he physical parameters 
  as output of the  simulation  on} NGC1265 }
 \label{fine_ngc1265}
 \[
 \begin{array}{ccc}
 \hline
 \hline
 \noalign{\smallskip}
 \mathrm {x_1~[{\mathrm{kpc}}]} & v_z~[{\mathrm{km/s}}] & \Delta~z
[{\mathrm{kpc}}] \\
 \noalign{\smallskip}
 \hline
 \noalign{\smallskip}
18.7 &  84.55 & 10.3 \\ \noalign{\smallskip}

\noalign{\smallskip}
 \hline
 \hline
 \end{array}
 \]

 \end {table}

\subsection{The first part of NGC4061}

\begin{figure}
  \begin{center}
\includegraphics[width=10cm]{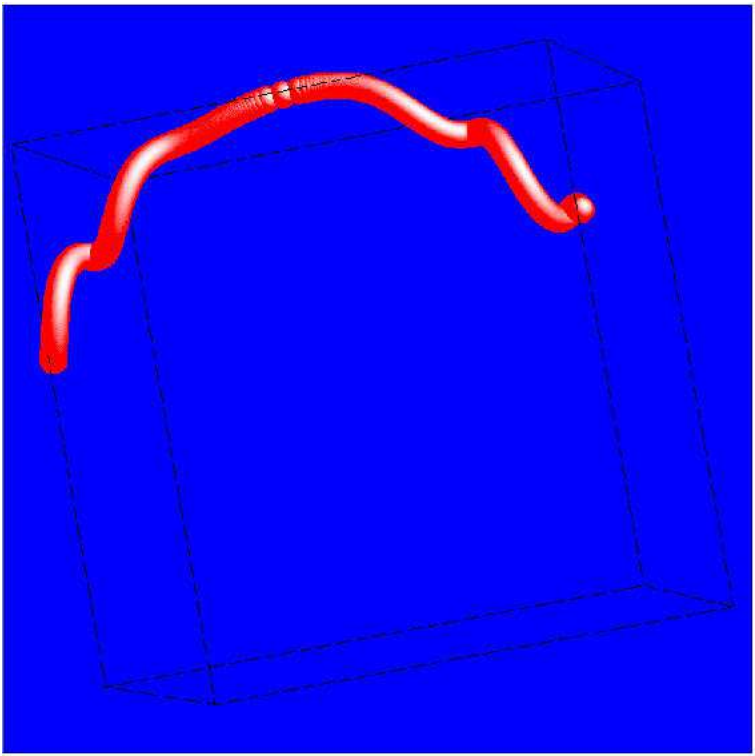}
\end {center}
\caption {
 Continuous three-dimensional trajectory of NGC4061:
 the three Eulerian angles (English notation)
 characterising the point of view are $ \Phi $=100   $^{\circ }$
 , $ \Theta $=-85   $^{\circ }$
 and  $ \Psi $=-10   $^{\circ }$.
 The precession is characterised by the angle
  $ \Psi_{prec} $=  5   $^{\circ }$
 and by the angular velocity
 $ \Omega_{prec} $= 60.00 [$^{\circ}/10^7 \mathrm{year}$].
 The three Eulerian angles are:
 $ \Phi_{j} $=  0   $^{\circ }$
 , $ \Psi_{j} $=  0   $^{\circ }$
 and  $ \Theta_{j} $= 90   $^{\circ }$.
 The angle of rotation of the galaxy is
 $ \alpha_{G} $=  0   $^{\circ }$.
 The physical parameters characterising the jet motion
 are : $E_{56}$=  0.02,
 $t_1$= 12.00~$10^7 \mathrm{year}$,
 $n_0$=  0.01~${\mathrm{particles/cm^3}}$,
 $x_1$= 20.00~${\mathrm{kpc}}$ ,
 $r_i$=  1.00~${\mathrm{kpc}}$ ,
 $\alpha_0$= 10   $^{\circ }$.
}
\label{NGC4061}%
    \end{figure}

\begin{figure}
  \begin{center}
\includegraphics[width=10cm]{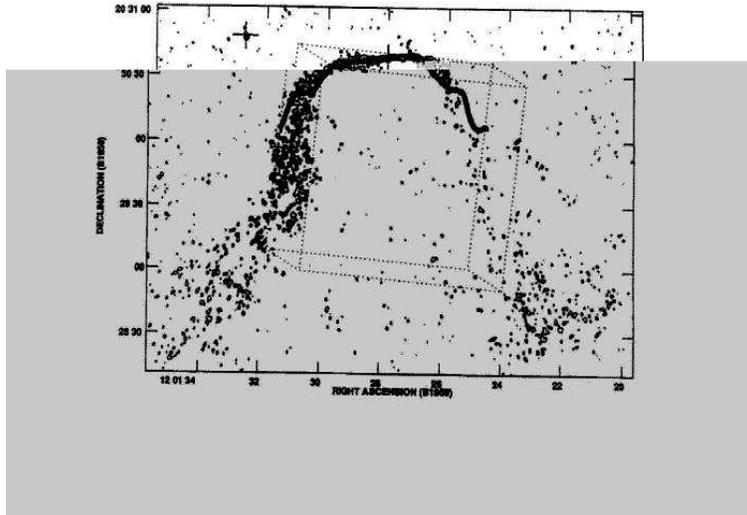}
  \end {center}
\caption {
Figure 5b of ~\citet{venkatesan} representing NGC4061 at 21cm superimposed on 
the theoretical trajectory represented by  Figure~\ref{NGC4061}.
}
\label{NGC4061_real}
\end{figure}

We have chosen the radio source NGC4061  at 21  cm,
see Figure 5b in 
~\citet{venkatesan}
 for a radio  map;
the basic data as extracted from the radio  observations
are reported in Table~\ref{data_ngc4061}.
The result of the simulation  is  visualised
in Figure~\ref{NGC4061}
and
in Figure~\ref{NGC4061_real} with the radio map 
superimposed on the simulation 
;  the caption of the Figure
shows  the adopted input parameters  while
the output data are reported  in
Table~\ref{fine_ngc4061}.
 \begin{table}
 \caption[]{T\lowercase{he data from the radio  map of} NGC4061 }
 \label{data_ngc4061}
 \[
 \begin{array}{cc}
 \hline
 \hline
 \noalign{\smallskip}
 \mathrm {L_{rad}[ergs/sec]} & mean ~spread~rate [degree] \\
 \noalign{\smallskip}
 \hline
 \noalign{\smallskip}
0.06 10^{40} &  10 \\ \noalign{\smallskip}
\noalign{\smallskip}
 \hline
 \hline
 \end{array}
 \]
 \end {table}

 \begin{table}
 \caption[]{T\lowercase{he physical parameters 
output of the simulation on} NGC4061 }
 \label{fine_ngc4061}
 \[
 \begin{array}{ccc}
 \hline
 \hline
 \noalign{\smallskip}
 \mathrm {x_1~[{\mathrm{kpc}}]} & v_z~[{\mathrm{km/s}}] & \Delta~z
[{\mathrm{kpc}}] \\
 \noalign{\smallskip}
 \hline
 \noalign{\smallskip}
20.9 &  78.65 & 14.6 \\ \noalign{\smallskip}
\noalign{\smallskip}
 \hline
 \hline
 \end{array}
 \]
 \end {table}

In this case we know that the dispersion velocity in the cluster
 is  485${\mathrm{km/sec}}$,
see  Table~1
in
~\citet{venkatesan}
.

\begin{figure}
  \begin{center}
\includegraphics[width=10cm]{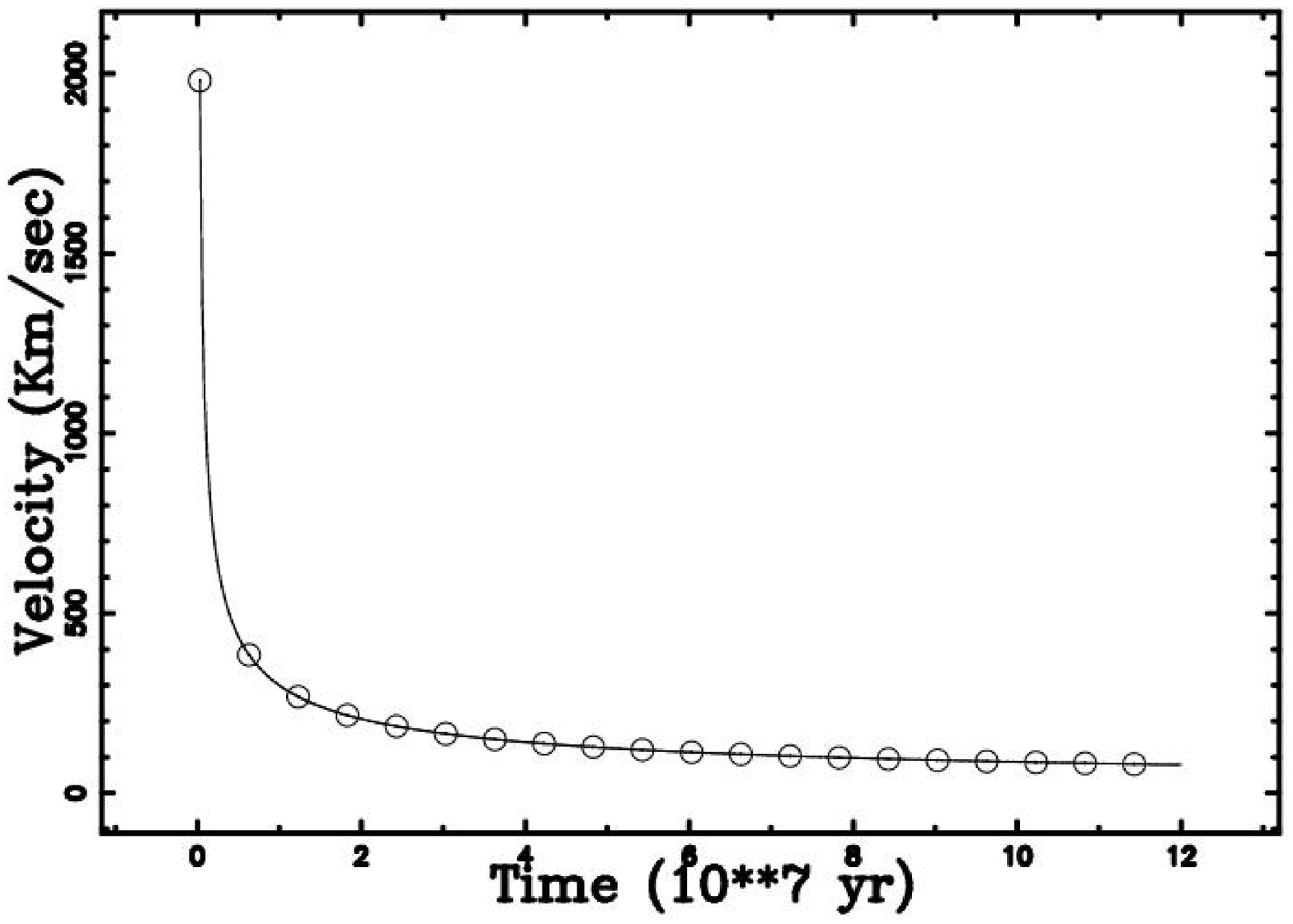}
\end {center}
\caption
{
The numerical relationship between velocity  and
$t_1$ in NGC4061. Parameters as in
Figure~\ref{NGC4061}.
}
\label{NGC4061_velo}%
    \end{figure}

The velocity $v_z$ in our model is  119.08${\mathrm{km/sec}}$
implying an angle of $\approx$ $14.2^{\circ}$ between
the direction of the galaxy motion and the
initial  direction of the jet.
A typical  plot  of the jet velocity  as  a function
of  the time is reported in Figure~\ref{NGC4061_velo}.

\subsection{The first part of NGC326}

\label{NGC326sec}
\begin{figure}
  \begin{center}
\includegraphics[width=10cm]{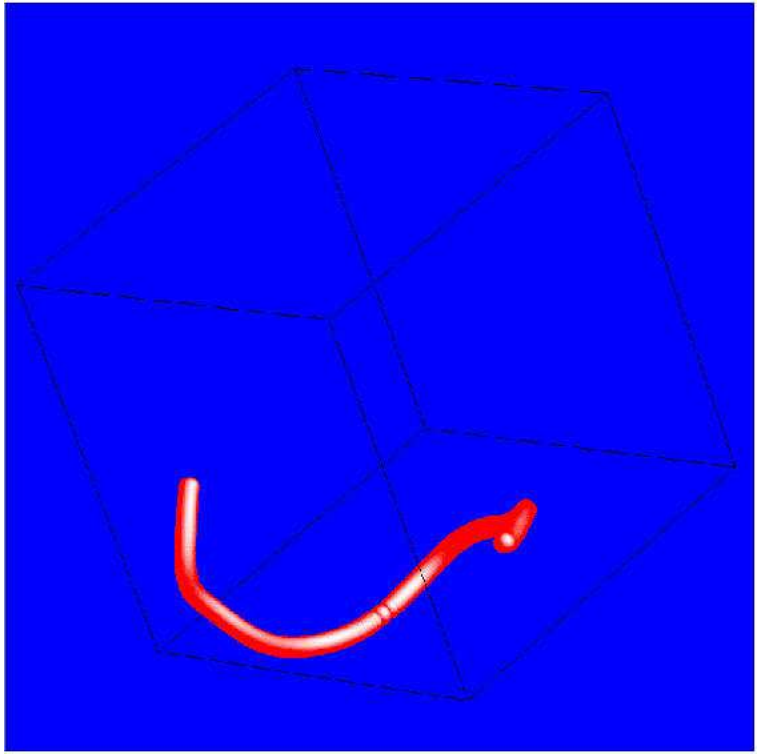}
\end {center}
\caption
{
 Continuous three-dimensional trajectory of NGC326:
 the three Eulerian angles (English notation)
 characterising the point of view are $ \Phi $=110   $^{\circ }$
 , $ \Theta $= 65   $^{\circ }$
 and  $ \Psi $=-40   $^{\circ }$.
 The precession is characterised by the angle
  $ \Psi_{prec} $=  2   $^{\circ }$
 and by the angular velocity
 $ \Omega_{prec} $= 51.43 [$^{\circ}/10^7\mathrm{year}$].
 The three Eulerian angles are:
 $ \Phi_{j} $=  0   $^{\circ }$
 , $ \Psi_{j} $=  0   $^{\circ }$
 and  $ \Theta_{j} $= 90   $^{\circ }$.
 The angle of rotation of the galaxy is
 $ \alpha_{G} $=-90   $^{\circ }$.
 The physical parameters characterising the jet motion
 are : $E_{56}$=  0.03,
 $t_1$= 14.00~$10^7$~$\mathrm{year}$,
 $n_0$=  0.01~${\mathrm{particles/cm^3}}$,
 $x_1$= 30.00~${\mathrm{kpc}}$ ,
 $r_i$=  1.50~${\mathrm{kpc}}$ ,
 $\alpha_0$=  6   $^{\circ }$.
}
\label{NGC326}%
    \end{figure}

\begin{figure}
  \begin{center}
\includegraphics[width=10cm]{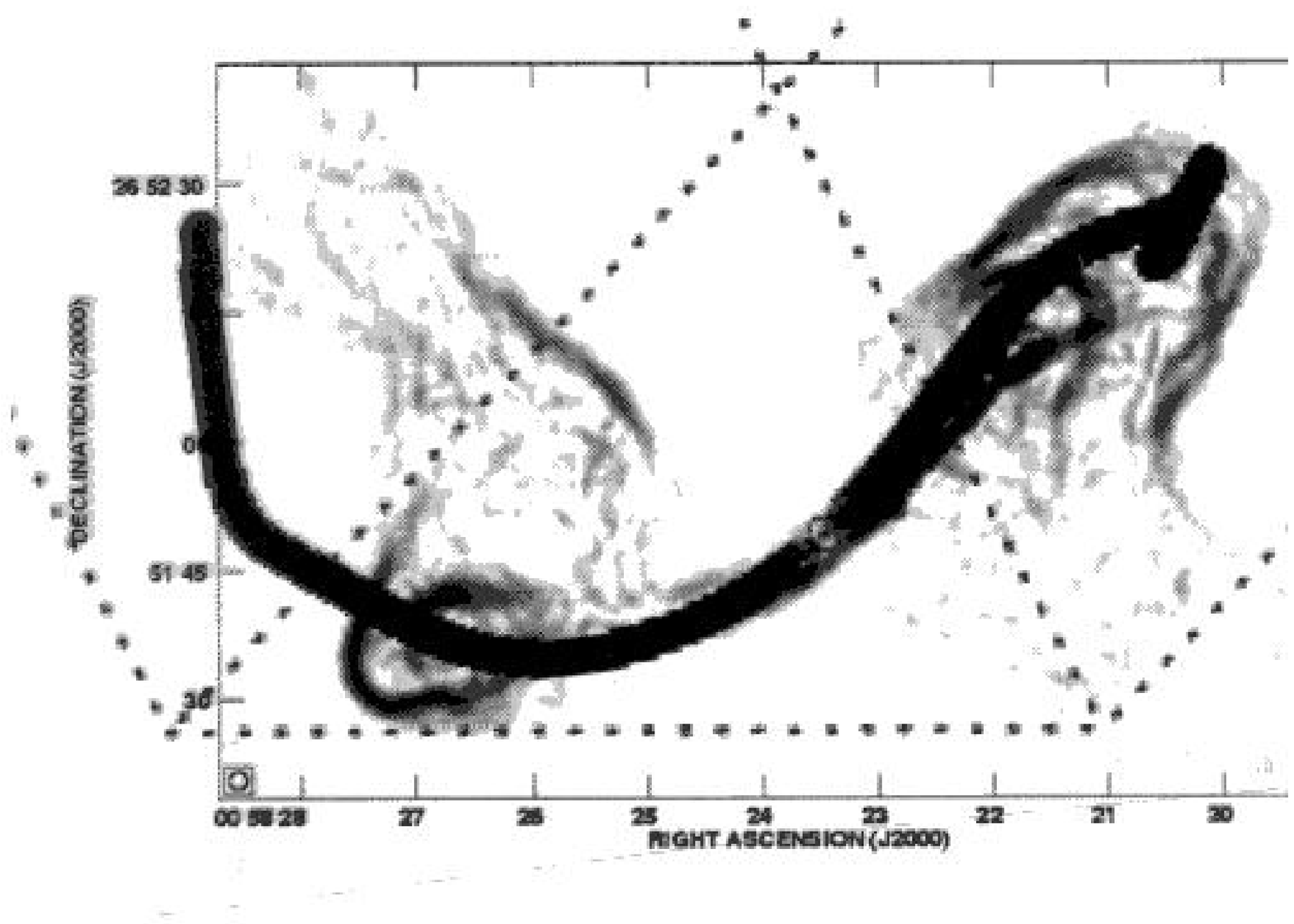}
  \end {center}
\caption {
Figure~6~(bottom panel) of ~~\citet{murgia}
superimposed on 
the theoretical trajectory represented by  Figure~\ref{NGC326}.
}
\label{NGC326_radio_theo}
\end{figure}

Another test of our code is the radio galaxy NGC326;
in particular
we refer to the image at  1.4GHz as given in
Figure~6~(bottom panel)
of
~\citet{murgia}
: the basic input data as deduced
from the radio  map are reported in Table~\ref{data_ngc326}.
 \begin{table}
 \caption[]{T\lowercase{he data from the radio  map of} NGC326  }
 \label{data_ngc326}
 \[
 \begin{array}{cc}
 \hline
 \hline
 \noalign{\smallskip}
 \mathrm {L_{rad}[ergs/sec]} & mean ~spread~rate [degree] \\
 \noalign{\smallskip}
 \hline
 \noalign{\smallskip}
0.5 10^{40} &  6.45 \\ \noalign{\smallskip}
\noalign{\smallskip}
 \hline
 \hline
 \end{array}
 \]
 \end {table}
In this case, the optical
observations
with HST reveal two cores, one coincident with the origin
of the radio galaxy, and the other  at the projected
distance of 4.8~${\mathrm{kpc}}$.
In this case,  it is  assumed that the rotation of the
host galaxy is due to
 gravitational interaction; the total angle of rotation
being  $\alpha_G $= $90~^\circ$.
 Figure~\ref{NGC326} reports on  the obtained simulation
and
 Figure~\ref{NGC326_radio_theo} reports  the radio map 
superimposed on the simulation ;
Table~\ref{fine_ngc326}  gives  the  data of output from
the simulation.

 \begin{table}
 \caption[]{T\lowercase{he physical parameters output of 
the simulation on } NGC326 }
 \label{fine_ngc326}
 \[
 \begin{array}{ccc}
 \hline
 \hline
 \noalign{\smallskip}
 \mathrm {x_1~[{\mathrm{kpc}}]} & v_z~[{\mathrm{km/s}}] & \Delta~z
[{\mathrm{kpc}}] \\
 \noalign{\smallskip}
 \hline
 \noalign{\smallskip}
30.6 &  98.71 & 16.8 \\ \noalign{\smallskip}
\noalign{\smallskip}
 \hline
 \hline
 \end{array}
 \]
 \end {table}

\subsection{A sample of WAT}
\label{secwat}
A sample of 7 radio--galaxies has already been presented in
Section~\ref{samplewat} and is now considered  as a test of the
theory  developed in Section~\ref{sezione_cont}. A first look at
the radio--maps  shows a nearly constant opening angle of
$\approx~5^{\circ}$. On assuming  that a constant ratio
 $E/x_1=E_{56}/1{\mathrm{kpc}}=10^{-3}$
characterises the motion,
  the terminal velocity $v_1$
 and   the age of the jets, $t_1$ ,
see  Table~\ref{sample},
can easily be found  through an iterative
procedure; the length of the jet
being provided in Table~7 of ~
~\citet{hardcastle}.
From  Table~\ref{sample}
the terminal velocities
can be compared
with the  central galaxy dispersion velocity
in the cluster.
 \begin{table}
 \caption[]{T\lowercase{he seven}  WAT \\
$\alpha_0$=$5^{\circ}$, \lowercase {$r_i[{\mathrm{kpc}} ]=x_1[{\mathrm{kpc}}]/100$,\\
~$n_0=10^{-3}{\mathrm{particles/cm^3}}$~and~}\\
       $E_{56}/1{\lowercase{\mathrm{kpc}}}=10^{-3}$ \\
        }
 \label{sample}
 \[
 \begin{array}{lccc}
 \hline
 \hline
 \noalign{\smallskip}
 Radio~name  & \mathrm {x_1~[{\mathrm{kpc}}]} & v_1~[{\mathrm{km/s}}] & t_1 [10^7\mathrm{
year}] \\
 \noalign{\smallskip}
 \hline
 \noalign{\smallskip}
0647+693 & 81 & 113  & 32  \\ \noalign{\smallskip}
1231+674 & 35 & 263  & 6.0  \\ \noalign{\smallskip}
1333+412 & 20 & 461  & 1.95  \\ \noalign{\smallskip}
1433+674 & 49 & 188  & 11     \\ \noalign{\smallskip}
2236-176 & 44 & 209  & 9.5     \\ \noalign{\smallskip}
3C465    & 28 & 329  & 3.8      \\ \noalign{\smallskip}
1610-608 & 13 & 708  & 0.82      \\ \noalign{\smallskip}
\noalign{\smallskip}
 \hline
 \hline
 \end{array}
 \]
 \end {table}
Most of the jets here considered are distinctly one sided over the length modelled,
and this fact is here justified by the flip-flop mechanism
rather than assuming that they are  relativistic over the whole
length.

\subsection{Power transition}

\label{sectransition} 
\begin{figure}
  \begin{center}
\includegraphics[width=10cm]{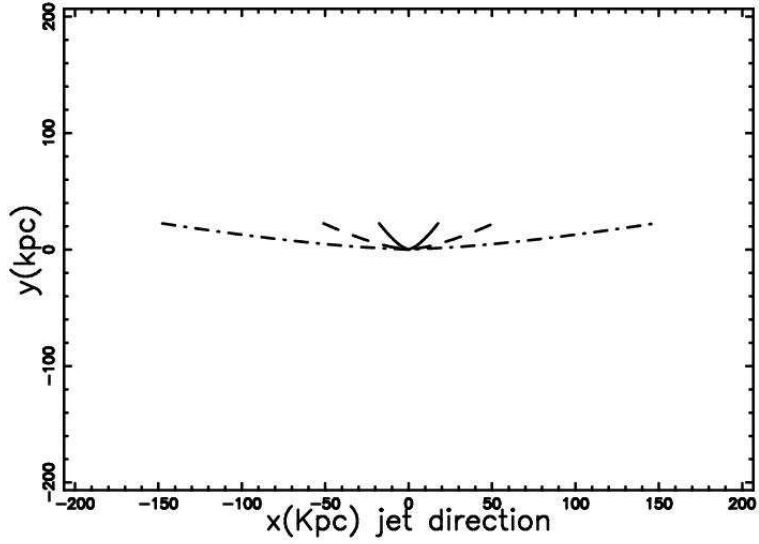}
\end {center}
\caption { Three  jet trajectories  as  a function of the power.
$P_{33}$=100     (dot-dash-dot-dash line) , $P_{33}$=1 (dashed
line    )  and $P_{33}$=0.01    (continuous  line). The other
parameters in common to the three trajectories are
$\alpha_0=5^{\circ}$, $n_0$=  0.001~${\mathrm{particles/cm^3}}$, $L_1$= 60
, $r_i$= $L_1$/20 , $t_7$=5.0 and $v_{400}$=1.
The three Eulerian angles are:
 $ \Phi_{j} $=  0   $^{\circ }$
 , $ \Psi_{j} $=  0   $^{\circ }$
 and  $ \Theta_{j} $= 90   $^{\circ }$.
The three Eulerian angles (English notation)
 characterising the point of view are $ \Phi $= 90   $^{\circ }$
 , $ \Theta $= 90   $^{\circ }$
 and  $ \Psi $= 0   $^{\circ }$.
}
\label{transition}
\end{figure}

The morphological  transition  between FR--I
and FR--II objects can be explained under the assumption of
constant power $P$, i.e. negligible synchrotron losses.
Equation~(\ref{eq:astror}) can be modified  by inserting the total power
$P=\frac{E}{t}$ and typical parameters of the jet previously
adopted such as   $\alpha_0=5^{\circ}$ and $n_0$=
0.001~${\mathrm{particles/cm^3}}$ :
\begin{equation}
x(t_7)=36.78\,{\mathrm{kpc}} \left( {\frac {{{\it t_7}}^{3}{\it P_{33}}\,
\sqrt [3]{{\it r_i}}}{{\it L_1}}} \right) ^{3/13} \quad ,
\label{eq:x}
\end {equation}
where $P_{33}=P/10^{33} Watt$.
Only
the velocity of the host galaxy in the direction
perpendicular to that of the jet  is  here considered
\begin{equation}
y (t_7) = 4.05 {\mathrm{kpc}} \; t_7\; v_{400} \quad  ,
 \label{eq:y}
\end{equation}
where $v_{400}=v[{\mathrm{km/sec}}]/400$.
When  a comparison of equation~(\ref{eq:x}) and (\ref{eq:y}) is made,
it is clear
that an increase in the mechanical power of the galaxy corresponds to an
increase in the elongation along x of the phenomena. A typical
plot that tentatively  reproduces the power transition is reported
in  Figure~\ref{transition}.

There are many 'straight' FR-I's, NGC315 is a good example .
This can  happen when the galaxy's velocity is lower
than the standard value here adopted of 400${\mathrm{km/s}}$  or 
from  a  particular  point of view of the observer.

\section{Application of the two--phase beam  }

\label{sezione_e} The two--phase  beam  developed in
Sections~\ref{twophase} and ~\ref{twophaserelativistic} is 
now applied to two radio-sources : in one  ,Cygnus~A,
 the relativistic--classic transition
is applied and in the other ,0647+693,  the classic--classic
transition.   

\subsection{Relativistic Cygnus~A}

A good  introduction to the classical
double radio sources ,
as well  as to Cygnus A,
 can be found in chapter 6 of
~\citet{deyoung}
.
The main physical parameters characterising the jet in
 Cygnus~A ,
as presented  in
~\citet{perley}
, are summarised  in
Table~\ref{data_cygnusA}; here $x_1^{obs}$
is the distance between nucleus and
hot-spots ,  $\alpha_0^{obs}$ is the first
opening angle (measured
on the radio map) and $t_{age}^{obs}$ , the age of the radio--galaxy,
as deduced
on the basis of the spectral steepening of the lobe emission.
 \begin{table}
 \caption[]{T\lowercase{he observed data on} Cygnus A  }
 \label{data_cygnusA}
 \[
 \begin{array}{cccc}
 \hline
 \hline
 \noalign{\smallskip}
  x_1^{obs}~[{\mathrm{kpc}}] & \alpha_0^{obs}~[^{\circ}]
         &t_{age}^{obs} [10^7 \mathrm{year}] \\
 \noalign{\smallskip}
 \hline
 \noalign{\smallskip}
  60 & 5 &   0.6 \\ \noalign{\smallskip}
\noalign{\smallskip}
 \hline
 \hline
 \end{array}
 \]
 \end {table}

\begin{figure}
  \begin{center}
\includegraphics[width=10cm]{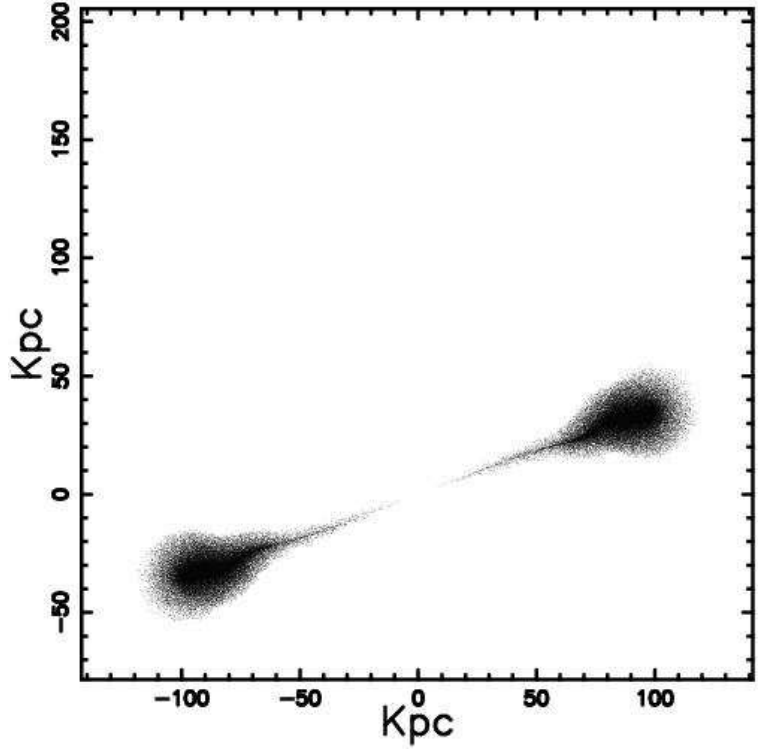}
\end {center}
\caption
{
 Projected 3D trajectory of CygnusA:
 the three Eulerian angles (English notation)
 characterising the point of view are $ \Phi $= 90   $^{\circ }$
 , $ \Theta $=  0   $^{\circ }$
 and  $ \Psi $=-20   $^{\circ }$.
 The precession is characterised by the angle
  $ \Psi_{prec} $=  1   $^{\circ }$
 and by the angular velocity
 $ \Omega_{prec} $=  600.00 [$^{\circ}/10^7$~$\mathrm{year}$].
 The three Eulerian angles are:
 $ \Phi_{j} $=  0   $^{\circ }$
 , $ \Psi_{j} $=  0   $^{\circ }$
 and  $ \Theta_{j} $= 90   $^{\circ }$.
 Parameters of the two--phase continuous model:
  $x_0$=    0.0~${\mathrm{kpc}}$,
  $r_0$=    0.02~${\mathrm{kpc}}$,
  $\beta_0 $=  0.8 ,
   $t_2$=  0.6~$10^7$~$\mathrm{year}$,
 $t_{1}$=  0.161~$10^7$~$\mathrm{year}$,
 $\alpha_0$ =  5   $^{\circ }$~
 and~ $\alpha_1$= 35   $^{\circ }$.
}
\label{CygnusA_proj}%
    \end{figure}

\begin{figure}
  \begin{center}
\includegraphics[width=10cm]{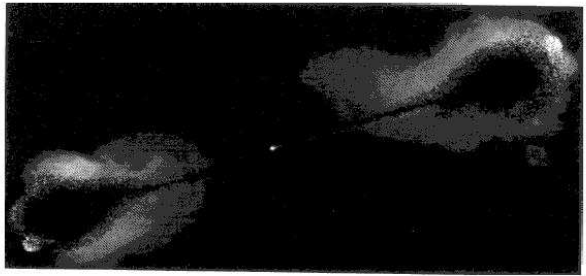}
  \end {center}
\caption {
Cygnus from the NRAO Image Gallery (white)
superimposed on 
the theoretical trajectory represented by  
Figure~\ref{CygnusA_proj} (black).
}
\label{CygnusA_radio_theo}
\end{figure}

Once the $t_2=t_{age}^{obs}$ is inserted in the code,
the other input parameters are varied
in order to obtain
  $x_1~\approx x_1^{obs}$;
this is obtained by solving  numerically the integral 
equation~(\ref{traiettoriarel}).
The results of the simulation (as  outlined in
Section~\ref{twophaserelativistic}) 
are reported in  Figure~\ref{CygnusA_proj}
and
in Figure~\ref{CygnusA_radio_theo} with the radio map 
superimposed on the simulation 
; Table~\ref{fine_cygnusA}  reports the data.

 \begin{table}
 \caption[]{T\lowercase{he physical parameters output of
the simulation on }Cygnus A }
 \label{fine_cygnusA}
 \[
 \begin{array}{cccccc}
 \hline
 \hline
 \noalign{\smallskip}
 physical~quantity\backslash index & 0 & 1 & 2 \\
 \noalign{\smallskip}
 \hline
 \noalign{\smallskip}
 x~[{\mathrm{kpc}}],distance  & 0.0     &  60.01 & 106.18    \\ \noalign{\smallskip}

 r~[\mathrm{kpc}],radius    & 0.02    &  5.25  & 19.81    \\ \noalign{\smallskip}

 v~[{\mathrm{km/s}}],velocity &  240000  &24577  & 6518    \\ \noalign{\smallskip}

 t~[10^7\mathrm{year}],time  &  0  &  0.161 & 0.6   \\ \noalign{\smallskip}

\noalign{\smallskip}
 \hline
 \hline
 \end{array}
 \]

 \end {table}

\subsection{Classic 0647+693}

A radio--map  of 0647+693 in the  8.5~GHz band
can be found in
~\citet{hardcastle}
where two radio--images were traced at the resolution
of $0.3\times0.6$ arcsec and 2 arcsec.
The influence of the motion of the host galaxy
is  evident from the maps and can be
visualised  in the following way
\begin{itemize}
\item  A line is traced between the  two hot--spots.
\item  The perpendicular to the previous line that intersects
       the nucleus  is traced.
       The distance along the perpendicular to the previous
       line turns out to be $\approx$~36.8~${\mathrm{kpc}}$.
\end {itemize}
\label{galvel}
From Figure~1 of ~\citet{hardcastle}
it is  also  possible to measure the  second opening angle ,
$\alpha_1$ that turns out to be  $\approx~25^{\circ}~$.

\begin{figure}
  \begin{center}
\includegraphics[width=10cm]{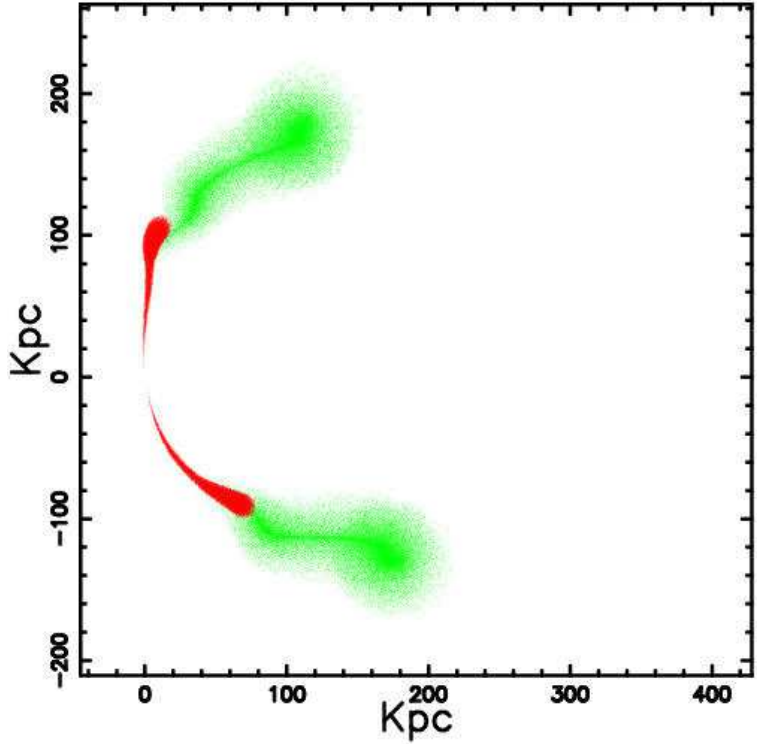}
\end {center}
\caption
{
 Projected 3D trajectory of 0647+693 (red points):
 the three Eulerian angles (English notation)
 characterising the point of view are $ \Phi $= 10   $^{\circ }$
 , $ \Theta $= 80   $^{\circ }$
 and  $ \Psi $=-10   $^{\circ }$.
 The precession is characterised by the angle
  $ \Psi_{prec} $=  5   $^{\circ }$
 and by the angular velocity
 $ \Omega_{prec} $=   22.50 [$^{\circ}/10^7\mathrm{year}$].
 The three Eulerian angles are:
 $ \Phi_{j} $=  0   $^{\circ }$
 , $ \Psi_{j} $=  0   $^{\circ }$
 and  $ \Theta_{j} $= 90   $^{\circ }$.
 Parameters of the two--phase continuous model:
 $x_0$=   0.0001~${\mathrm{kpc}}$,
  $r_0$=   0.0600~${\mathrm{kpc}}$,
  $c_f$=      5.0,
   $t_2$=  8.00~$10^7\mathrm{year}$,
 $t_{1}$=  4.00~$10^7\mathrm{year}$,
 $\alpha_0$ =  5   $^{\circ }$~
 and~ $\alpha_1$= 25   $^{\circ }$. 
The green points represents the same trajectory after
$t_2$=  32.00~$10^7{\mathrm{year}}$.}
\label{0647_proj}%
    \end{figure}

The results of the simulation (as  outlined in
Section~\ref{twophase}) are reported in  Figure~\ref{0647_proj}(red points)
and
in Figure~\ref{0647_radio_theo_1}, with the radio map 
superimposed on the simulation
; Table~\ref{fine_0647} reports the data.
\begin{figure}
  \begin{center}
\includegraphics[width=10cm]{Figure14.eps}
  \end {center}
\caption {
Figure~2 of ~\citet{hardcastle}
superimposed on 
the theoretical trajectory represented by  Figure~\ref{0647_proj}(red
points).}
\label{0647_radio_theo_1}
\end{figure}

 \begin{table}
 \caption[]{T\lowercase{he physical parameters output of 
 the simulation on} 0647+693 }
 \label{fine_0647}
 \[
 \begin{array}{cccccc}
 \hline
 \hline
 \noalign{\smallskip}
 physical~quantity\backslash position & 0 & 1 & 2 \\
 \noalign{\smallskip}
 \hline
 \noalign{\smallskip}
 x~[{\mathrm{kpc}}],distance  & 0.0001  &  80.7 & 104.4    \\ \noalign{\smallskip}

 r~[{\mathrm{kpc}}],radius  & 0.060  &  3.6 & 8.8    \\ \noalign{\smallskip}

 v~[{\mathrm{km/sec}}],velocity &  60000.0 & 1004.7 & 407.6   \\ \noalign{\smallskip}

 t~[10^7~\mathrm{year}],time  &  0  &  4.0 & 8.0   \\ \noalign{\smallskip}

\noalign{\smallskip}
 \hline
 \hline
 \end{array}
 \]
 \end {table}

\begin{figure}
  \begin{center}
\includegraphics[width=10cm]{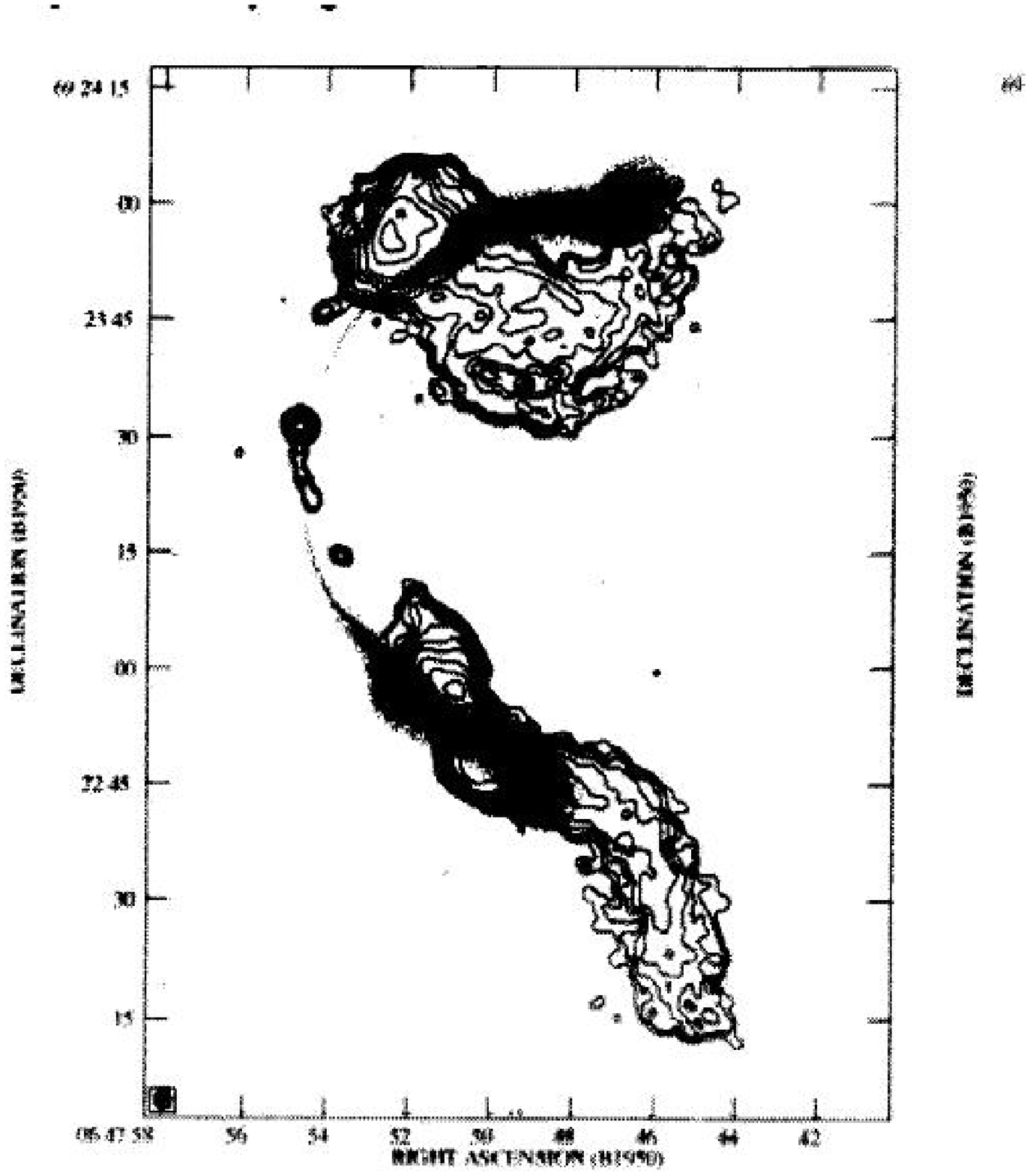}
  \end {center}
\caption {
Figure~1 of ~\citet{hardcastle}
superimposed on 
the theoretical trajectory represented by  Figure~\ref{0647_proj}(green
points).
}
\label{0647_radio_theo_2}
\end{figure}

In order to reproduce the long trails of  0647+693
visible in Figure~1 of~\citet{hardcastle},
the simulation is now performed
at  a value of $t_1$ which is bigger by
factor four in respect to the
previous run ,
see Figure~\ref{0647_proj}(green points)
and
Figure~\ref{0647_radio_theo_2},  where the radio map 
is superimposed on the simulation.

\section{The  Kelvin-Helmholtz  instabilities}
\label{kh}

 The very small opening angles of  observed extra-galactic
radio--jets ,  if interpreted as free expansion,  in many cases
imply a Mach number  $M$ $>$ 6 and existing theories are briefly
reviewed.
Some basic questions can be posed on the stability of
such highly supersonic jets :
\begin{itemize}
\item   Is the jet unstable against  perturbation
        with  a wavelength smaller than the radius?

\item   Is the jet stable over long distances?

\item   If the jet is unstable,  how does the distance
        over which an
        instability grows by a factor of e
        depend upon M and the density
        contrast between the jet and the external medium?

\item   Are  the helical instabilities with an azimuthal number
        $m$ = 1 and  the pinch instabilities , $m$=0 ,
        enough to build a theory on  the lateral
        growth of  the jet due to an increase
        of  energy  in the perturbations ?
\end {itemize}

In order to study these questions,  we must analyse
  the
Kelvin-Helmholtz instability of an axisymmetric flow along the
$z$-axis when the wavelengths $\lambda=\frac{2 \pi}{k} $ ( $k$ is the
wave-vector) are smaller or  greater than   the jet radius $a$,
which is taken to be independent of the position along the  jet.
The velocity , $U_0$, is assumed to be rectangular. The internal (
external ) fluid density is represented by $\rho_{0i}$
($\rho_{0e}$)   , the internal sound velocity is $s$ and  $\nu_0$=
$\frac {\rho_{0i}}{\rho_{0e}}$. The analysis is  therefore split
in two.

\subsection{Small wavelengths}

If the growth rate
of the
envelope of the reflecting modes is analysed,
it is possible to deduce the following approximate
relationship for the imaginary ($\gamma_{KH}$)
and real ($\omega_r$)   part of
the perturbations when $M~>~1$
\begin {equation}
\gamma_{KH} =  \gamma_{ad}  \frac{s}{a}
\end   {equation}
with
\begin{equation}
\gamma_{ad}  \approx 0.17 (\ln (ka M))^{1.21} ( \frac {2}{\nu_0 +
1/\nu_0})^{0.06} (ka)^{0.7} \quad  ,
\end{equation}
and
\begin {equation}
\omega_r  \approx  \frac{M}{2} \nu_0 ^{0.21} ka \frac {s}{a}
\quad .
\end {equation}
More details can be found in~\citet{zaninetti1986}.

\subsection{Great  wavelengths}

Starting from the equations of motion
and   continuity,  and
assuming both fluids to be adiabatically compressible,
it
is possible to derive and to solve
the dispersion relation numerically , see \citet{zaninetti1987}.

We then start from observable  quantities
that can be  measured  on radio-maps
such as   the total length $L_{obs}$,
the wavelength   $\lambda_{obs}^1$  of the wiggles ($m$=1)
along the jet,
the distance $\lambda_{obs}^0$ ($m$=0)  between knots,
and the final offset
$\Delta\,L_{obs}$  of the center of the jet.

These observable quantities  are  identified
with   the following theoretical  variables:
\begin {equation}
  \lambda^1_{max} = \lambda^1_{obs}
\quad,
\end   {equation}
\begin {equation}
  \lambda^0_{max} = \lambda^0_{obs}
\quad,
\end   {equation}
\begin {equation}
  A_0 \exp \bigl (\frac {2L_{obs}} {M t_{ad} a}) =  \Delta L_{obs}
\quad,
\end   {equation}
\begin {equation}
 nl_e  = L_{obs}
\quad,
\end   {equation}
where $t_{ad}$ = $t_{min } \cdot s_i/a $
and $A_0$ is the amplitude of the
perturbed energy.
The result is  a theoretical  expression   for  $t_{min}$
the  minimum time scale of the instability,
$\lambda_{max}$  the wavelength connected with  the most
unstable  mode and  $l_e$   the distance over which the most
unstable  mode grows  by a factor  $e$,
see ~\citet{zaninetti1987}.
These  parameters  can then be found
through the  set  of nonlinear equations
previously  reported.
By choosing  three  objects,
M87~(~\citet{owen1980})
, NGC6251~(see ~\citet{perley1984b})
and  NGC1265~( see ~\citet{odea})
the   observational  parameters can be measured
on the radio image, see Table~\ref{observ}.
 \begin{table}
 \caption[]{P\lowercase{arameters of the observed oscillations in radio--galaxies
expressed in averaged radius units} }
 \label{observ}
 \[
 \begin{array}{lccc}
 \hline
 \hline
 \noalign{\smallskip}
Geometrical~misuration                   &  M87   & NGC6251 & NGC1265 \\
 \noalign{\smallskip}
 \hline
 \noalign{\smallskip}
L_{obs}        [averaged~radius~units]  & 48    & 58.75 & 20   \\ \noalign{\smallskip}
\lambda^0_{obs}[averaged~radius~units]  & 4.0   & 4.89  & 4  \\    \noalign{\smallskip}
\lambda^1_{obs}[averaged~radius~units]  &  19.2 & 17.5  & 13.3  \\  \noalign{\smallskip}
\Delta L_{obs} [averaged~radius~units]  &  0.96 & 5.87  & 2  \\  \noalign{\smallskip}
\noalign{\smallskip}
 \hline
 \hline
 \end{array}
 \]
 \end {table}

The four nonlinear equations are then solved
and  the  four theoretical parameters
are found , see Table~\ref{theor}.
 \begin{table}
 \caption[]{T\lowercase{heoretical parameters from  oscillations deduced
from the four nonlinear equations} }
 \label{theor}
 \[
 \begin{array}{lccc}
 \hline
 \hline
 \noalign{\smallskip}
Theoretical~variable         &  M87    & NGC6251 &      NGC1265        \\
 \noalign{\smallskip}
 \hline
 \noalign{\smallskip}
n                              & 2.75   & 2.88   & 1.1               \\ \noalign{\smallskip}
M                              & 17.28  & 19.44  & 17.93               \\ \noalign{\smallskip}
\nu_0                          & 9.61   & 28.05  & 9.4               \\ \noalign{\smallskip}
A_0                            & 5.57~10^{-3}  & 8.2~10^{-3} & 0.25  \\ \noalign{\smallskip}
 \hline
 \hline
 \end{array}
 \]
 \end {table}

\begin{figure}
  \begin{center}
\includegraphics[width=10cm]{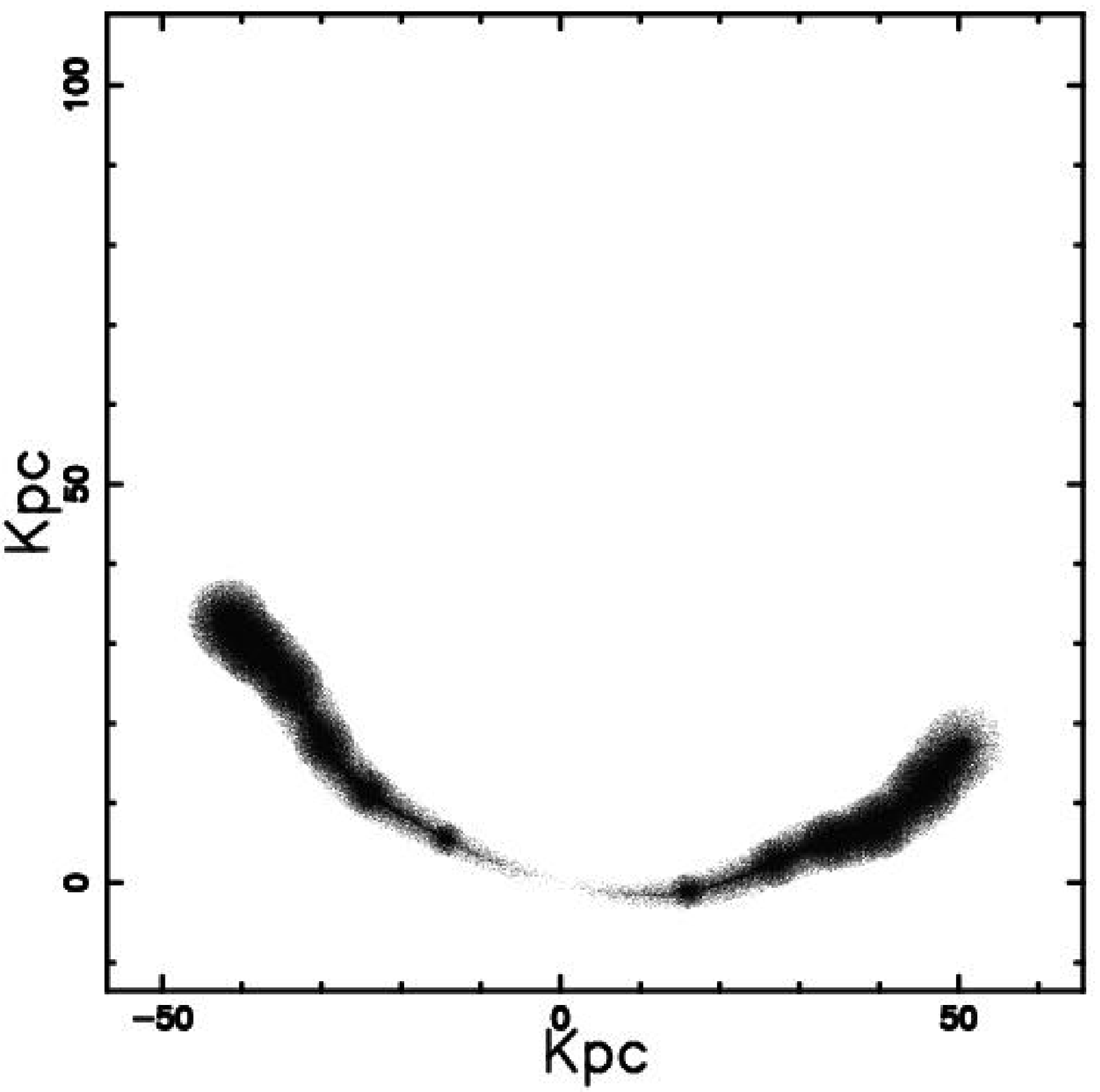}
\end {center}
\caption
{
Parameters as in Figure~\ref{NGC1265_proj},
but with
superimposition of the pinch mode ($m=0$).
}
\label{NGC1265_kh}
\end{figure}

\begin{figure}
  \begin{center}
\includegraphics[width=10cm]{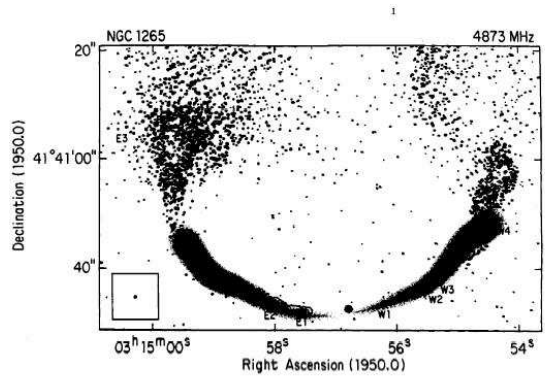}
  \end {center}
\caption {
Figure 1 of ~\citet{odea} representing NG1265 at 4873 MHz superimposed on 
the pinch mode represented by  Figure~\ref{NGC1265_kh}.
}
\label{NGC1265_kh_theo}
\end{figure}

An application of the  results here obtained  is  reported
in Figure~\ref{NGC1265_kh}
where   the wavelength
of the pinch modes ($m=0$)
and the oscillations of the helical mode ($m=1$)
as reported in Table~\ref{theor}
are  simulated by identifying the lateral growth
of energy  with  the precession.
Figure~\ref{NGC1265_kh_theo} reports  the radio map 
superimposed on the simulation of the pinch modes.

\section{Summary}

\subsection{New Physics}
The concept of deceleration is
both  an explanation of observed properties
of a sample of radio-galaxies
,see~\citet{laing1999} ,
as well as   an hypothesis of work to explain
the deceleration  of a relativistic jet of electrons
and positron pairs  by mass injection, see~\citet{laing2002} ,
or by adiabatic models that require
coupling between the variations of velocity,
magnetic field and particle density,
see~\citet{laing2002}.

Here two new laws of motion for extra-galactic jets
in the  presence of laminar flow are derived ,
equation~(\ref{first}) and equation~(\ref{second}),
once the two canonical laws of resistance are introduced.

The introduction of the momentum conservation
along a solid angle
that characterises the extra-galactic
jet allows us to find:
\begin {itemize}
\item
An analytical  law 
,equation~(\ref{third}),
for the velocity and  length of the jet
as a  function
of the other structural parameters which  are:
the involved linear density of energy,
the initial radius,
the elapsed time and the solid angle.
This is the third law of motion.
\item
An analytical law  for the velocity if conversely
the conservation of the
momentum's flux in a turbulent  jet is assumed, 
equation~(\ref{traiettoria}).
This is the fourth law of motion.
\item
An  integral equation~(\ref{traiettoriarel})
of motion if 
the conservation of the relativistic
momentum's flux in a turbulent  jet is assumed.
This is the fifth law of motion.
\end {itemize}
The new equation~(\ref{astroviscosity}) connects the turbulent eddy viscosity
with
the jet's opening angle.

\subsection{Morphological Results}

The introduction of the law of motion
allows us to find
a space trajectory on introducing
the jet precession,
the velocity and rotation
of the
galaxy,
and the inclination of the jet with respect to the plane
of the galaxy.

It can therefore be suggested   that
the hydrodynamical effects necessary to bend the jet (see
for example Section 4.4.5 of \citet{deyoung} or the discussion in
\citet{odea1985}) 
can also be explained as an artifact of 
a change in the system
of reference combined with precession effects 
of the jet as well
as decreasing jet velocity.
From an observational point of view it is not easy to discriminate 
between the bending of the radio-galaxy due to the dynamic 
pressure , see equation~(A35) in \citet{deyoung} and
the bending due to the decreasing jet's velocity.
Both models require a virial speed of the host galaxy $\approx$
1000${\mathrm{\frac{km}{s}}}$ and differ only in the requirement
for  the jet's velocity which  is constant in the traditional
approach and variable in the models here adopted.
This dilemma will be  solved
when  measurements of the jet's velocity independent from
the radio-models will be available.
The spectroscopic observations for radio galaxies having ultra
steep radio spectra, for example ,
show 
velocity shifts range from 100 to almost 1000 $\mathrm{km/s}$ 
, see~\citet{Miley1997}.

\subsection{Other approaches }

In FR-I jets, which are  thought to decelerate, the profile of this
deceleration appears in  other models 
that are now briefly reviewed 
\begin{itemize}
\item In \citet {blk1996}  
 a detailed study of the dynamical effects of entraining cool
(thermally sub-relativistic) material into hot' (thermally relativistic)
jets  is  presented. The dissipation associated with entrainment 
causes only modest loss of kinetic energy flux, and it is
 shown that relativistic jets are affected much
less by dissipation than are classical flows.
Their equation (41)  represents the flux  of kinetic energy 
but no analytical results are given for the relativistic law of motion
such as  our formula (\ref{traiettoriarel}).
\item In \citet {lb2002}   
the  conservation of particles, energy and momentum  are applied 
in order to derive the
variations of pressure and density along
the jets of the low-luminosity radio galaxy 3C 31.
Their 
self-consistent solutions for deceleration by injection of thermal matter
can be  compared with our  equation~(\ref{third}).
\item  In \citet{Canvin_Laing}
some  functional forms  are introduced  to describe the
geometry, velocity, emissivity and magnetic-field structure  
of the two low-luminosity radio galaxies B2 0326+39 and B2 1553+24
and  an accurate comparison between models and data is presented.
Their solutions are purely numerical and 
analytical solutions such as our formula (\ref{traiettoriarel}) are absent.
\end{itemize}

\subsection{Analytical results and  hydrodynamical codes}

A comparison with  the hydrodynamical codes is tentatively
made  on the following key--points
\begin{itemize}
\item  The hydrodynamical codes are usually performed on a family
of parameters given by $M_b$ , which is  the beam Mach number , 
while $\rho_b$ is the
inside/outside density contrast and t  is the time over which the
phenomena is followed. Due to the fact that the hydrodynamical
codes do  not give the possibility of  closing
 the equations with
parameters deduced from the observations,  $\infty$ figures
should be produced in order to find the  one that better
approximates the radiosource chosen to be simulated, see for
example~\citet{norman1996} and
        \citet{balsara1992}.
Conversely,
the analytical approach  developed here , see 
 eqns.~(\ref{first}),(\ref{second}),
(\ref{eq:astror}) and (\ref{eq:astrov}),
 gives position and velocity once a few input parameters
are provided. 
\item The treatment here adopted can not follow  the
nonlinear development of the K-H instabilities in 3D   , see for
example~\citet{hardee} where the simulation was  performed at
$M_b$=5; but despite  this inconvenience, the concept of developed
turbulence with its  consequent constant density  along the jet is
widely  used, see  Section~\ref{sezione_cont} and
Section~\ref{twophase}.
\end {itemize}
\subsection{Average density}
Here we have assumed that the density of 
the jet is constant
and equal to its surrounding medium.  
This is certainly something
that is not necessarily the case for many 
astrophysical jets.
Some  properties can be deduced on the medium surrounding 
the elliptical 
galaxies from the X-ray surface  brightness  distribution.
Assuming  that the gas is isothermal,
 the following  empirical  
law 
(~\citet{Fabbiano1989}) is   used for the 
electron density:
\begin {equation} 
n_e(r) =  n_e(0)  
\left [  1  + ( \frac {r}{ax})^2 \right ]^{-3 \beta /2}
\quad , 
\label{elliptical}
\end {equation}
here $\beta$ $\sim$ 0.4  $\div$ 0.6 , $n_e(0)$ $\sim$ 0.1
$\mathrm{particles} \over {\mathrm{cm}^3}$  and  
ax $\sim$  1$\div$ 3${\mathrm{ kpc}}$.  
On  adopting  this radial dependence 
for proton density we can  
evaluate the average density 
when $n_e(0) = 0.1 \frac {\mathrm{particles} } {\mathrm{cm}^3}$,
ax = 2.0 ${\mathrm{kpc}}$   and  $\beta$= 0.4.
The   average  value  of
$n_0$   ,  $\overline{n_0}$ , when evaluated over  10000 points
in the interval  0-100 ${\mathrm{kpc}}$
is    $\overline{n_0}$ =
$6.74\;10^{-3} \frac {\mathrm{particles} } {\mathrm{cm}^3}$.
This  is the theoretical value  of $n_0$ that should
be used , for example , in equation~(\ref{eq:astror}).

\appendix
\section{The roots of a cubic }
\label{appendix_cubic}
To solve the cubic polynomial equation 
\begin{equation}
a_{{0}}{x}^{3}+a_{{1}}{x}^{2}+a_{{2}}x+a_{{3}}
=0
\quad ,
\label{cubicequation}
\end{equation}
 for $x$,
the first step is to apply the  transformation
\begin{equation}
x=y-1/3\,{\frac {a_{{1}}}{a_{{0}}}}
\quad .
\label{transxy}
\end{equation}
This reduces the equation to 
\begin{equation}
y^3 + py + q = 0
\quad ,
\label{cubicequation_reduced}
\end{equation}
where
\begin{eqnarray*}
p & = & 1/3\,{\frac {3\,a_{{2}}a_{{0}}-{a_{{1}}}^{2}}{{a_{{0}}}^{2}}}     \\
q & = &1/27\,{\frac
{27\,a_{{3}}{a_{{0}}}^{2}+2\,{a_{{1}}}^{3}-9\,a_{{2}}a_{{
1}}a_{{0}}}{{a_{{0}}}^{3}}} . \\
\end{eqnarray*}
The next step is to 
compute the first derivative  of  the  left hand side
of equation~(\ref{cubicequation_reduced}) calling it $f(y)^{\prime}$
\begin{equation}
f(y)^{\prime}= 3y^2 + p  
\quad .
\label{firstderivate}
\end{equation}
In the case in which  $p \geq 0 $  ,
in the range of existence   $- \infty  < y  < \infty $  , 
the first derivative is always   positive 
and equation~(\ref{cubicequation_reduced})
has  only one   root that is real , more precisely ,
\begin{equation}
y=1/6\,\sqrt [3]{-108\,{\it q}+12\,\sqrt {12\,{{\it p}}^{3}+81\,{{\it 
q}}^{2}}}-2\,{\frac {{\it p}}{\sqrt [3]{-108\,{\it q}+12\,\sqrt {12
\,{{\it p}}^{3}+81\,{{\it q}}^{2}}}}}
\quad ,
\label{pqsolution}
\end{equation}
or equation~(\ref{cubicequation}) has the solution
in terms of $a_0$,$a_1$,$a_2$ and $a_3$
\begin{eqnarray}
1/6\,\sqrt [3]{-36\,{\frac {3\,a_{{2}}a_{{0}}-{a_{{1}}}^{2}}{{a_{{0}}}
^{2}}}+12\,\sqrt {4/9\,{\frac { \left( 3\,a_{{2}}a_{{0}}-{a_{{1}}}^{2}
 \right) ^{3}}{{a_{{0}}}^{6}}}+9\,{\frac { \left( 3\,a_{{2}}a_{{0}}-{a
_{{1}}}^{2} \right) ^{2}}{{a_{{0}}}^{4}}}}} \nonumber \\
-2/3\, \left( 3\,a_{{2}}a_{
{0}}-{a_{{1}}}^{2} \right) {a_{{0}}}^{-2}{\frac {1}{\sqrt [3]{-36\,{
\frac {3\,a_{{2}}a_{{0}}-{a_{{1}}}^{2}}{{a_{{0}}}^{2}}}+12\,\sqrt {4/9
\,{\frac { \left( 3\,a_{{2}}a_{{0}}-{a_{{1}}}^{2} \right) ^{3}}{{a_{{0
}}}^{6}}}+9\,{\frac { \left( 3\,a_{{2}}a_{{0}}-{a_{{1}}}^{2} \right) ^
{2}}{{a_{{0}}}^{4}}}}}}} . \\
\label{asolution}
\end{eqnarray}

\end{document}